\documentclass[12pt]{article}
\usepackage{plain}
\pdfoutput=1

\usepackage{chemformula}
\usepackage[version=4]{mhchem}
\usepackage{xstring}
\usepackage{xcolor}
\usepackage{ifthen}
\usepackage{graphicx}

\usepackage[colorlinks]{hyperref}
\hypersetup{
    colorlinks=true,
    citecolor = olive,
    linkcolor=blue, 
    linktoc=none
}

\usepackage{amssymb}
\usepackage{amsmath}
\usepackage{amsfonts}
\usepackage{epsfig}
\usepackage{anysize}
\usepackage[right]{lineno}

\marginsize{2cm}{2cm}{2.5cm}{3cm}

\date{}

\def\bs{\boldsymbol}     

\makeatletter   
  \newcommand*\bigcdot{\mathpalette\bigcdot@{.5}}
  \newcommand*\bigcdot@[2]{\mathbin{\vcenter{\hbox{\scalebox{#2}{$\m@th#1\bullet$}}}}}
\makeatother

\usepackage{tocloft}

\makeatletter
  \renewcommand \thesection {\@Roman\c@section}  
  \renewcommand\thesubsection {\thesection.\@roman\c@subsection}
  \newcommand\myss{\thesection-\@roman\c@subsection} 
  \renewcommand\thesubsubsection {\thesubsection.\@arabic\c@subsubsection}
  \newcommand\mysss{\myss-\@arabic\c@subsubsection} 
  \renewcommand\subsubsection{\@startsection {subsubsection}{3}{\z@ }
    {-1.5ex\@plus -1ex \@minus -.2ex}
    {0.8ex \@minus 0ex\nointerlineskip\vspace{-0.5\lineskip}}  
    {\normalfont \footnotesize \bfseries }}
\makeatother

\usepackage[compact]{titlesec}          
\AtBeginDocument{
  \setlength\abovedisplayskip{40pt}
  \setlength\belowdisplayskip{0pt}}

\titlespacing{\section}
     {0pt}{6.5ex plus 1ex minus .2ex}{1.3ex plus .2ex minus .2ex} 

\usepackage[finalnew]{trackchanges}

\addeditor{r1}
\addeditor{CV}

\def\nPub(#1){{\footnotesize\noindent
\ifcase #1 ()
    {}            
  \or
    \clasif{\bf aa 1.}
        {Random-walk theory and Ornstein-Zernike systems with extended-core  potentials.}
        {\AUT{A. Robledo, I.E. Farquhar}  
        \emph{J. Chem. Phys.} \textbf{61}, 1594 (1974)}   {\url{https://doi.org/10.1063/1.1682143}}
  \or
    \clasif{\bf aa 2.}
        {Random-walk theory and correlation functions in classical statistical mechanics.}
	{\AUT{A. Robledo, I.E. Farquhar}  
          \emph{Physica A} \textbf{84}, 435 (1976)}
        \url{https://doi.org/10.1016/0378-4371(76)90097-2}
  \or
    \clasif{\bf aa 3.}
        {Random-walk theory, ordered phases in lattice gas systems.}
	{\AUT{A. Robledo, I.E. Farquhar}  
          \emph{Physica A} \textbf{84}, 449 (1976)}
        \url{https://doi.org/10.1016/0378-4371(76)90098-4}
  \or
    \clasif{\bf aa 4.}
        {Random-walk theory and the decay of pair correlations in Ornstein-Zernike lattice systems.}
	{\AUT{A. Robledo, I.E. Farquhar}  
          \emph{Physica A} \textbf{84}, 472 (1976)}
        \url{https://doi.org/10.1016/0378-4371(76)90099-6}
  \or
    \clasif{\bf aa 5.}
        {On the relationship between continuous time random walks and the nonequilibrium Ornstein-Zernike equation.}
	{\AUT{A.B. Budgor, A. Robledo}  
          \emph{Physica A} \textbf{85}, 329 (1976)}
        \url{https://doi.org/10.1016/0378-4371(76)90053-4}
  \or
    \clasif{\bf aa 6.}
        {Random walks, the Ornstein-Zernike equation and the condensation of a 1-dimensional lattice gas.}
	{\AUT{A. Robledo, A.B. Budgor}  
          \emph{Am. J. Phys.} \textbf{46}, 998 (1978)}
        \url{https://doi.org/10.1119/1.11491}
  \or
    \clasif{\bf aa 7.}
        {Multiple trapping of random walkers on periodic space lattices.}
	{\AUT{A. Robledo, L. Woodhouse}  
          \emph{J. Stat. Phys.} \textbf{19}, 129 (1978)}
        \url{https://doi.org/10.1007/BF01012507}
  \or
    \clasif{\bf aa 8.}
        {Surface electronic Green's functions in terms of the bulk Green's function.}
	{\AUT{C. Varea, A. Robledo}  
          \emph{Phys. Rev. B} \textbf{19}, 1310 (1979)}
        \url{https://doi.org/10.1103/PhysRevB.19.1310}
  \or
    \clasif{\bf aa 9.}
        {Pair correlations for a hard-core lattice gas against a hard wall.}
	{\AUT{A. Robledo}  
        \emph{Mol. Phys.} \textbf{39}, 193 (1980)}
  \or
    \clasif{\bf aa 10.}
        {Tight binding Green's functions for surfaces, thin  films  and  solid 	interfaces. A random  walk approach.}
	{\AUT{A. Robledo, C. Varea}  
          \emph{Phys. Rev. B} \textbf{21}, 1469 (1980)}
        \url{https://doi.org/10.1103/PhysRevB.21.1469}
  \or
    \clasif{\bf aa 11.}
        {The liquid-solid transition of the hard sphere system from uniformity of the 	chemical potential.}
	{\AUT{A. Robledo}  
        \emph{J. Chem. Phys.} \textbf{72}, 1701 (1980)}   {\url{https://doi.org/10.1063/1.439281}}
  \or
    \clasif{\bf aa 12.}
        {Rigorous interface properties of the van der Waals mixture.}
	{\AUT{A. Robledo, A. Valderrama, C. Varea}  
          \emph{J. Chem. Phys.} \textbf{73}, 6365 (1980)}
        \url{https://doi.org/10.1063/1.440123}
  \or
    \clasif{\bf aa 13.}
        {On the relationship between the density functional formalism and the potential distribution theory for nonuniform systems.}
	{\AUT{A. Robledo, C. Varea}  
          \emph{J. Stat. Phys.} \textbf{26}, 513 (1981)}
        \url{https://doi.org/10.1007/BF01011432}
  \or
    \clasif{\bf aa 14.}
        {Nucleation, spinodal decomposition and kinetics of phase change in the	van der Waals fluid.}
	{\AUT{C. Varea, A. Robledo}  
          \emph{J.\ Chem.\ Phys.} \textbf{75}, 5080 (1981)}
        \url{https://doi.org/10.1063/1.441899}
  \or
    \clasif{\bf aa 15.}
        {Kinetics of phase change in a model binary alloy.}
	{\AUT{A. Robledo, C. Varea}  
          \emph{Phys. Rev. B} \textbf{25}, 4711, (1982)}
        \url{https://doi.org/10.1103/PhysRevB.25.4711}
  \or
    \clasif{\bf ma 16.}
        {Free energy density functionals for nonuniform classical fluids.
          In: Keller, J., G\'azquez, J.L. (eds) Density Functional Theory.}
	{\AUT{A. Robledo, C. Varea}  
          \emph{Lecture Notes in Physics} \textbf{187}, 287 (1983)}
        \url{https://doi.org/10.1007/3-540-12721-6_10}
  \or
    \clasif{\bf m 17.}
       {Nonuniform relaxation of thermodynamic fluctuations and kinetics of phase change.}
	{\AUT{A. Robledo, C. Varea}  
        Memorias de la 1a Escuela Mexicana de F\'{\i}sica Estad\'{\i}stica
         	(Soc. Mex. F\'{\i}s., 1983)}
  \or
    \clasif{\bf aa 18.}
        {Global phase diagram for the wetting transition at interfaces in fluid mixtures.}
	{\AUT{M.E. Costas, C. Varea, A. Robledo}  
        \emph{Phys. Rev. Lett.} \textbf{51}, 2394 (1983)}
        \url{https://doi.org/10.1103/PhysRevLett.51.2394}
  \or
    \clasif{\bf aa 19.}
        {Prewetting in partially miscible liquids and the structure and	thermodynamics of transient foams and aerosols.}
	{\AUT{J. Gracia, C. Varea, A. Robledo}  
          \emph{J. Phys. Chem. (Lett.)} \textbf{88}, 3923 (1984)}
        \url{https://doi.org/10.1021/j150662a004}
  \or
    \clasif{\bf m 20.}
       {Wetting, prewetting, foams and emulsions.}
	{\AUT{A. Robledo, C. Varea, J. Gracia, C. Guerrero, J. Lla\~nes, G. Irazoque}  
        \emph{Kinam} \textbf{6A}, 47 (1984)}
  \or
    \clasif{\bf m 21.}
       {Wetting at a Donnan membrane.}
	{\AUT{E. Martina, C. Varea, A. Robledo}  
        \emph{Kinam} \textbf{6A}, 57 (1984)}
  \or
    \clasif{\bf m 22.}
       {Statistical mechanics of nonuniform systems. Applications to interfaces.}
	{\AUT{A. Robledo, C. Varea}  
         Memorias de la 2a Escuela Mexicana de F\'{\i}sica Estad\'{\i}stica
         	(Soc. Mex. F\'{\i}s., 1984)}
  \or
    \clasif{\bf e 23.}
       {Termodin\'amica, Unidad V,Laboratorio de Ciencias B\'asicas I.}
	{\AUT{A. Robledo}  
        Facultad de Qu\'{\i}mica, UNAM, 1984}
  \or
    \clasif{\bf e 24.}
       {Naturaleza de las transiciones de mojado en intercaras de mezclas fluidas binarias.}
	{\AUT{C. Varea, M.E. Costas, A. Robledo}  
        Cuadernos de Posgrado (de la FQUNAM) 15, 93 (1984)}
  \or
    \clasif{\bf d 25.}
       {Las transiciones interfaciales.}
	{\AUT{A. Robledo}  
        \emph{Ciencia} \textbf{35}, 125 (1984)}
  \or
    \clasif{\bf aa 26.}
        {Wetting regimes at semipermeable membranes.}
	{\AUT{C. Varea, A. Robledo, E. Martina}  
          \emph{Phys. Rev. A} \textbf{31}, 1825 (1985)}
        \url{https://doi.org/10.1103/PhysRevA.31.1825}
  \or
    \clasif{\bf aa 27.}
        {Interfacial critical phenomena at semipermeable membranes.}
	{\AUT{A. Robledo, C. Varea, E. Martina}  
          \emph{Phys. Rev. B} \textbf{32}, 7545 (1985)}
        \url{https://doi.org/10.1103/PhysRevB.32.7545}
  \or
    \clasif{\bf aa 28.}
        {Pinning of antiphase boundaries at the cleaved (001) surface of an L$2{_0}$ ordering alloy .}
	{\AUT{L. Vicente, C. Varea, A. Robledo}  
          \emph{Surf. Sci.} \textbf{164}, 479 (1985)}
        \url{https://doi.org/10.1016/0039-6028(85)90761-7}
  \or
    \clasif{\bf aa 29.}
        {Roughening transition and formation of bicontinuous structures of immiscible solvents embedded in surfactant diblock copolymers.}
	{\AUT{A. Robledo, C. Varea, Martina, E.}  
          \emph{J. Phys. Lett. (fr)} \textbf{46}, L-967 (1985)}
        \url{https://doi.org/10.1051/jphyslet:019850046020096700}
  \or
    \clasif{\bf m 30.}
       {The wetting transition.}
	{\AUT{C. Varea, A. Robledo}  
        Memorias de la VI Reuni\'on de Bajas Temperaturas
        (J. Heiras, T. Akachi and R. Tsumura, editores, IIMUNAM, 1985)}
  \or
    \clasif{\bf aa 31.}
        {Relationships between the phase behavior of lattice models of amphiphile mixtures
           and Griffiths's Three-Component Model.}
	{\AUT{C. Varea, A. Robledo}  
          \emph{Phys. Rev. A} \textbf{33}, 2760 (1986)}
        \url{https://doi.org/10.1103/PhysRevA.33.2760}
  \or
    \clasif{\bf aa 32.}
        {Transient foaminess, micelle formation and wetting behavior in water-phenol mixtures.}
	{\AUT{J. Gracia, C. Guerrero, J. Lla\~nes, A. Robledo}  
          \emph{J. Phys. Chem.} \textbf{90}, 1350 (1986)}
        \url{https://doi.org/10.1021/j100398a028}
  \or
    \clasif{\bf aa 33.}
        {Exact thermodynamic correspondence between a lattice model microemulsion and simpler spin systems .}
	{\AUT{A. Robledo}  
          \emph{EPL} \textbf{1}, 303 (1986)}
        \url{http://iopscience.iop.org/0295-5075/1/6/006}
  \or
    \clasif{\bf aa 34.}
        {The distribution of hard rods on a line of finite length.}
	{\AUT{A. Robledo, J.S. Rowlinson}  
          \emph{Mol. Phys.} \textbf{58}, 711 (1986)}
        \url{https://doi.org/10.1080/00268978600101521}
  \or
    \clasif{\bf ma 35.}
        {General Discussion remarks on wetting  transitions, confined fluids and microemulsion models.}
	{\AUT{A. Robledo}  
        \emph{J. Chem. Soc. Faraday Trans. II} \textbf{82}, 1841-1844, 1848, 1855 (1986)}
  \or
    \clasif{\bf e 36.}
       {La transici\'on nem\'atico-isotr\'opico.}
	{\AUT{M.E. Costas, C. Varea, A. Robledo}  
        \emph{Cuadernos de Posgrado (de la FQUNAM)} \textbf{23}, 17 (1986)}
  \or
    \clasif{\bf aa 37.}
        {A nonlocal theory of irreversible thermodynamics.}
	{\AUT{A. Robledo, C. Varea}  
          \emph{J. Noneq. Thermodyn.} \textbf{12}, 213 (1987)}
        \url{https://doi.org/10.1515/jnet.1987.12.3.213} 
  \or
    \clasif{\bf aa 38.}
        {Critical magnetization at antiphase boundaries of magnetic binary alloys.}
	{\AUT{C. Varea, A. Robledo}  
          \emph{Phys. Rev. B} \textbf{36}, 5561 (1987)}
        \url{https://doi.org/10.1103/PhysRevB.36.5561}
  \or
    \clasif{\bf aa 39.}
        {Spin Ising transcription of a lattice model of micellar solutions.}
	{\AUT{A. Robledo}  
          \emph{Phys. Rev. A} \textbf{36}, 4067 (1987)}
        \url{https://doi.org/10.1103/PhysRevA.36.4067}
  \or
    \clasif{\bf ma 40.}
        {Oxygen ordering and twin boundaries in a Landau-Ginsburg superconductor oxide model.}
	{\AUT{C. Varea, A. Robledo}  
        ``Novel Superconductivity'', S.A. Wolf and V.Z. Kresin, Eds., Plenum, 1987, pp. 1033-1039}
  \or
    \clasif{\bf aa 41.}
        {Effect of oxygen pressure and temperature on the tetragonal-orthorhombic transition in a model $\ce{YBa2Cu3O_{7-y}}$.}
	{\AUT{C. Varea, A. Robledo}  
          \emph{Rev. Mex. F\'{\i}s.} \textbf{33}, 311 (1987)}
        \url{https://rmf.smf.mx/ojs/index.php/rmf/article/view/1936}        
  \or
    \clasif{\bf m 42.}
       {Statistical mechanics of micellar solutions and microemulsions.}
	{\AUT{A. Robledo}  
        \emph{Kinam} \textbf{8A}, 5 (1987)}
  \or
    \clasif{\bf aa 43.}
        {High-Tc superconductivity at twin boundaries in a Landau-Ginzburg oxide model.}
	{\AUT{A. Robledo, C. Varea}  
          \emph{Phys. Rev. B} \textbf{37}, 637 (1988)}
        \url{https://doi.org/10.1103/PhysRevB.37.631}
  \or
    \clasif{\bf ma 44.}
        {Oxygen absorption and structural transition in a model $\ce{YBa2Cu3O_{7-y}}$.} 
	{\AUT{C. Varea, A. Robledo}  
          \emph{MRS Symp. Proc.} \textbf{99}, 527 (1988)}
        \url{https://doi.org/10.1557/PROC-99-527}
  \or
    \clasif{\bf ma 45.}
        {On the formation of twins in Orthorhombic $\ce{YBa2Cu3O_{7-y}}$.}
	{\AUT{D. Rios-Jara, C. Varea, A. Robledo, A. Huanosta, J.M. Dominguez, T. Akachi, R. Escudero}  
        \emph{MRS Symp. Proc.} \textbf{99}, 233 (1988)}
  \or
    \clasif{\bf aa 46.}
        {Model for oxygen absorption and structural phase transition in $\ce{YBa2Cu3O_{7-y}}$.}
	{\AUT{C. Varea, A. Robledo}  
          \emph{Mod. Phys. Lett. B} \textbf{2}, 1017, (1988)}
        \url{https://doi.org/10.1142/S0217984988000849}
  \or
    \clasif{\bf ma 47.}
        {Crystalline structure in the $\ce{Cu}$ basal plane and microestructure in superconducting $\ce{YBa2Cu3O_{7-y}}$.}
	{\AUT{C. Varea, A. Robledo}  
        Progress in High Temperature Superconductivity, Vol. 5, p.145
        (World scientific Publishing Co. 1988)}
  \or
    \clasif{\bf aa 48.}
        {Continuous wetting transition at the liquid-vapor interface of the binary liquid mixture cyclohexane-acetonitrile.}
	{\AUT{L.M. Trejo, J. Gracia, C. Varea, A. Robledo}  
          \emph{EPL} \textbf{7}, 537 (1988)}
        \url{https://doi.org/10.1209/0295-5075/7/6/010}
  \or
    \clasif{\bf aa 49.}
        {Simple spin-hole model for magnetic correlations in copper-oxide superconductors.}
	{\AUT{A. Robledo, C. Varea}  
          \emph{Int. J. Mod. Phys. B} \textbf{1}, 763 (1988)}
        \url{https://doi.org/10.1142/S0217979288000597}
  \or
    \clasif{\bf e 50.}
       {Statistical mechanical models for micellar solutions and microemulsions.}
	{\AUT{A. Robledo}  
        Proceedings of the Fourth Mexican School on Statistical Physics, World 	Scientific, Singapore, 1988, pp. 93-165}
  \or
    \clasif{\bf e 51.}
       {Landau theory for pure interfacial transitions.}
	{\AUT{C. Varea, A. Robledo}  
        Proceedings of the Fourth Mexican School on Statistical Physics, World 	Scientific, Singapore, 1988, pp. 215-233}
  \or
    \clasif{\bf d 52.}
       {Interfacial phase transitions.}
	{\AUT{A. Robledo}  
         Advanced Series in Surf. Sci., 2, p.251,
      	World Scientific, Singapore, 1988,}
  \or
    \clasif{\bf aa 53.}
        {Sublattice ordered phases of the Griffiths' Three-Component model.}
	{\AUT{V. Talanquer, C. Varea, A. Robledo}  
          \emph{Phys. Rev. B} \textbf{39}, 7016 (1989)}
        \url{https://doi.org/10.1103/PhysRevB.39.7016}
  \or
    \clasif{\bf aa 54.}
        {Global phase diagram for binary alloys with one magnetic component.}
	{\AUT{V. Talanquer, C. Varea, A. Robledo}  
          \emph{Phys. Rev. B} \textbf{39}, 7030 (1989)}
        \url{https://doi.org/10.1103/PhysRevB.39.7030}
  \or
    \clasif{\bf aa 55.}
        {Sublattice-ordered phases in a model for a micellar solution.}
	{\AUT{V. Talanquer, C. Varea, A. Robledo}  
          \emph{Phys. Rev. B} \textbf{39}, 7039 (1989)}
        \url{https://doi.org/10.1103/PhysRevB.39.7039}
  \or
    \clasif{\bf aa 56.}
        {Spin-hole model for magnetic phase diagram and pairing mechanism in copper-oxide superconductors.}
	{\AUT{A. Robledo, C. Varea}  
          \emph{Rev. Mex. F\'{\i}s.} \textbf{35}, 255 (1989)}
        \url{https://rmf.smf.mx/ojs/index.php/rmf/article/view/2050}
  \or
    \clasif{\bf ma 57.}
        {New source of corrections to scaling for micellar solution critical behavior.}
	{\AUT{G. Martinez-Mekler, G.F. Al-Noaimi, A. Robledo}  
        NATO Advanced Study Institute Series B, Pergamon Press, 1989, pp. 211-215}
  \or
    \clasif{\bf ma 58.}
        {Magnetic vortex-antivortex pairing mechanism for doped $\ce{La2CuO4}$.}
	{\AUT{A. Robledo, C. Varea}  
          \emph{Physica C} \textbf{162-164}, 1517 (1989)}
        \url{https://doi.org/10.1016/0921-4534(89)90800-9}
  \or
    \clasif{\bf ma 59.}
        {Global behavior of the diffusion coefficient for the van der Waals binary mixture.}
	{\AUT{R. Castillo, M.E. Costas, A. Robledo}  
          \emph{Int. J. Thermophys.} \textbf{10}, 427 (1989)}
        \url{https://doi.org/10.1007/BF01133539}
  \or
    \clasif{\bf ma 60.}
        {Spin-hole model for magnetism and superconductivity in copper-oxide supercounductors.}
	{\AUT{A. Robledo, C. Varea}  
        \emph{Cond. Matt. Theories, Vol.} \textbf{4}, p.281, Plenum Press, 1989.}
  \or
  \clasif{\bf m 61.}
        {The hard-sphere order-disorder transition in the Bethe continuum.}
	{\AUT{A. Robledo}  
        Lecture Notes in Thermodynamics and Statistical Mechanics
	A.E. Gonzalez, C. Varea and M. Medina-Noyola, eds. p. 52, World Scientific, 1989.}
  \or
    \clasif{\bf m 62.}
       {Mean-field properties of lattice models for micellar solutions.}
	{\AUT{V. Talanquer, C. Varea, A. Robledo}  
        Lecture Notes in Thermodynamics and Statistical Mechanics
      	A.E. Gonzalez, C. Varea and M. Medina-Noyola, eds. p. 64, World Scientific, 1989.}
  \or
    \clasif{\bf aa 63.}
        {Uncommon source of corrections to scaling for micellar solution critical behavior.}
	{\AUT{G. Martinez-Mekler, G.F. Al-Noaimi, A. Robledo}  
          \emph{Phys. Rev. A} \textbf{41}, 4513 (1990)}
        \url{https://doi.org/10.1103/PhysRevA.41.4513}
  \or
    \clasif{\bf aa 64.}
        {Spin-hole model with magnetic vortex-antivortex pairing mechanism in copper-oxide superconductors.}
	{\AUT{A. Robledo, C. Varea}  
          \emph{Physica C} \textbf{166}, 334 (1990)}
        \url{https://doi.org/10.1016/0921-4534(90)90414-A}
  \or
    \clasif{\bf ma 65.}
        {Spin-hole model with magnetic vortex-antivortex pairing mechanism for 	doped $\ce{La2CuO4}$ .}
	{\AUT{A. Robledo, C. Varea}  
        ``Oxygen Disorder Effects in High Tc Superconductors'', p. 169
       	Plenum Press, 1990}
  \or
    \clasif{\bf m 66.}
       {Curvature interfacial transitions at amphiphile monolayers and their
        possible relation to the onset of micelle formation.}
	{\AUT{A. Robledo, C. Varea, V. Talanquer}  
        Lecture Notes in Thermodynamics and Statistical Mechanics
       	M. Lopez de Haro and C. Varea, eds. p. 3, World Scientific, 1990.}
  \or
    \clasif{\bf m 67.}
       {Micellar solution model with asymmetric solubility loops and anomalous
        critical isochore with crossover from angular to parallel approach to coexistence.}
        {\AUT{A. Robledo, G. Martinez-Mekler, C. Varea}  
	Lecture Notes in Thermodynamics and Statistical Mechanics
        	M. Lopez de Haro and C. Varea, eds. p. 14, World Scientific, 1990.}
  \or
    \clasif{\bf e 68.}
       {Soluciones micelares and microemulsiones.}
	{\AUT{A. Robledo}  
        Escuela de Verano: La Visi\'on Molecular de la Materia
	L. Moch\'an e I. Ortega, eds.
       	Universidad Aut\'onoma del Estado de Morelos, pp. 609-646, 1990}
  \or
    \clasif{\bf d 69.}
       {Microemulsiones: estructura and aplicaciones.}
	{\AUT{A. Robledo, G. Ruiz}  
        Ciencias, Revista de Difusi\'on, FCUNAM, octubre 1990. p. 18}
  \or
    \clasif{\bf aa 70.}
        {Curvature interfacial transitions in amphiphile monolayers and their possible relation to the onset of micelle formation.}
	{\AUT{A. Robledo, C. Varea, V. Talanquer}  
          \emph{Phys. Rev. A} \textbf{43}, 5736 (1991).}
        \url{https://doi.org/10.1103/PhysRevA.43.5736}
  \or
    \clasif{\bf aa 71.}
        {Asymmetric solubility loops and anomalous geometry at the lower critical point in a model micellar solution.}
	{\AUT{A. Robledo, G. Martinez-Mekler, C. Varea}  
          \emph{EPL} \textbf{16}, 405 (1991)}
        \url{https://doi.org/10.1209/0295-5075/16/4/015}
  \or
    \clasif{\bf aa 72.}
        {The hard-sphere order-disorder transition in the Bethe continuum.}
	{\AUT{A. Robledo, C. Varea}  
          \emph{J. Stat. Phys.} \textbf{63}, 1163 (1991)}
        \url{https://doi.org/10.1007/BF01030004}
  \or
    \clasif{\bf ma 73.}
        {Density functional theory beyond the surface tension. Interfacial width, elastic bending constants and line tension.}
	{\AUT{A. Robledo, C. Varea, V. Romero-Roch\'{\i}n}  
          \emph{Physica A} \textbf{177}, 474 (1991)}
        \url{https://doi.org/10.1016/0378-4371(91)90189-J}
  \or
    \clasif{\bf aa 74.}
        {Microscopic expressions for interfacial bending constants and spontaneous curvature.}
	{\AUT{V. Romero-Roch\'{\i}n, C. Varea, A. Robledo}  
          \emph{Phys. Rev. A} \textbf{44}, 8417 (1991)}
        \url{https://doi.org/10.1103/PhysRevA.44.8417}
  \or
    \clasif{\bf m 75.}
       {Bending rigidity of the liquid-vapor interface.}
	{\AUT{V. Romero-Roch\'{\i}n, C. Varea, A. Robledo}  
        Lectures on Thermodynamics and Statistical Mechanics, World Scientific, Singapore, pp. 10-25, 1991.}
  \or
    \clasif{\bf m 76.}
       {Statistical mechanics of the line tension.}
	{\AUT{C.\ Varea, A.\ Robledo}  
        Lectures on Thermodynamics and Statistical Mechanics, World Scientific, Singapore, pp.\ 190-203, 1991.}
  \or
    \clasif{\bf aa 77.}
        {Evidence for the divergence of the line tension at the wetting transition.}
	{\AUT{C. Varea, A. Robledo}  
          \emph{Phys. Rev. A} \textbf{45}, 2645 (1992)}
        \url{https://doi.org/10.1103/PhysRevA.45.2645}
  \or
    \clasif{\bf aa 78.}
        {Wetting transition for the contact line and Antonov's rule for the line tension.}
	{\AUT{A. Robledo, C. Varea, J.O. Indekeu}  
          \emph{Phys. Rev. A} \textbf{45}, 2423 (1992)}
        \url{https://doi.org/10.1103/PhysRevA.45.2423}
  \or
    \clasif{\bf aa 79.}
        {Statistical Mechanics of the line tension.}
	{\AUT{C.\ Varea, A.\ Robledo}  
          \emph{Physica A} \textbf{183}, 12 (1992)}
        \url{https://doi.org/10.1016/0378-4371(92)90175-P}
  \or
    \clasif{\bf aa 80.}
        {Extended capillary-wave theory for the liquid-vapor interface and its width in the limit of vanishing gravity.}
	{\AUT{V. Romero-Roch\'{\i}n, C. Varea, A. Robledo}  
          \emph{Physica A} \textbf{184}, 367 (1992)}
        \url{https://doi.org/10.1016/0378-4371(92)90312-E}
  \or
    \clasif{\bf m 81.}
       {Interfacial phase transitions underlying amphiphile micellar self-assembly.}
	{\AUT{A. Robledo, C. Varea}  
        Lectures on Thermodynamics and Statistical Mechanics, World Scientific, Singapore, pp. 37-43, 1992.}
  \or
    \clasif{\bf ma 82.}
        {Novel interfacial shape transitions in amphiphile monolayers.}
	{\AUT{A. Robledo, C. Varea}  
        Cond. Matt. Theories, Vol. 8, Plenum Press, Nueva York, p. 595, 1992.}
  \or
    \clasif{\bf aa 83.}
        {Magnitude of the prewetting boundary tension near wetting for short-range forces.}
	{\AUT{C. Varea, A. Robledo}  
          \emph{Phys. Rev. E} \textbf{47}, 3772 (1993)}
        \url{https://doi.org/10.1103/PhysRevE.47.3772}
  \or
    \clasif{\bf aa 84.}
        {Hyperscaling and nonclassical exponents for the line tension at wetting.}
	{\AUT{J.O. Indekeu, A. Robledo}  
          \emph{Phys. Rev. E} \textbf{47}, 4607 (1993)}
        \url{https://doi.org/10.1103/PhysRevE.47.4607}
  \or
    \clasif{\bf aa 85.}
        {Strongly coupled Ising chain under a weak random field.}
	{\AUT{P.A. Mello, A. Robledo}  
        \emph{Physica A} \textbf{199}, 363 (1993)}
  \or
    \clasif{\bf aa 86.}
        {Stress tensor of curved interfaces.}
	{\AUT{V. Romero-Roch\'{\i}n, C. Varea, A. Robledo}  
          \emph{Mol. Phys.} \textbf{80}, 821 (1993)}
        \url{https:/doi.org/10.1080/00268979300102681}
  \or
    \clasif{\bf aa 87.}
        {Universality and the contact line at first-order wetting transitions.}
	{\AUT{A. Robledo, J.O. Indekeu}  
          \emph{EPL} \textbf{25}, 17 (1994).}
        \url{https:/doi.org/10.1209/0295-5075/25/1/004}
  \or
    \clasif{\bf aa 88.}
        {Magnetic anisotropy and superconductivity in a model for high-$T_c$ copper oxides.}
	{\AUT{A. Robledo}  
          \emph{Physica C} \textbf{220}, 271 (1994).}
        \url{https://doi.org/10.1016/0921-4534(94)90913-X}
  \or
    \clasif{\bf ma 89.}
        {Effect of thermal fluctuations on the singular behaviour of the line tension at wetting.}
	{\AUT{J.O. Indekeu, A. Robledo}  
        Proceedings of ``Recent advances in Statistical Physics'',
        \emph{Turkish J. Phys.} \textbf{18}, 285 (1994).}
  \or
    \clasif{\bf m 90.}
       {First-order wetting transitions as interface critical points.}
	{\AUT{A. Robledo, J.O. Indekeu}  
        XXIII Winter Meeting on Statistical Physics, Lectures on Thermodynamics and Statistical Mechanics,
        World Scientific, Singapore, p. 120, 1994.}
  \or
    \clasif{\bf aa 91.}
        {Free energy expressions for a spherical interface.}
	{\AUT{C. Varea, A. Robledo}  
          \emph{Mol. Phys.} \textbf{85}, 477 (1995).}
        \url{https://doi.org/10.1080/00268979500101261}
  \or
    \clasif{\bf aa 92.}
        {Scaling properties of the capillary-wave model with interfacial bending rigidity.}
	{\AUT{A. Robledo, C. Varea}  
       	  \emph{Mol. Phys.} \textbf{86}, 879 (1995)}
        \url{https://doi.org/10.1080/00268979500102451}
  \or
    \clasif{\bf ma 93.}
        {Bending rigidities and spontaneous curvature of spherical interfaces.}
        {\AUT{C. Varea, A. Robledo}  
        Proceedings of 24rd Winter Meeting on Statistical Physics, 3-6 January 1994, Cuernavaca, Morelos, Mexico, 
        \emph{Physica A} \textbf{220}, 33-47 (1995).}
  \or
    \clasif{\bf ma 94.}
        {The capillary-wave model with interfacial bending rigidity and its scaling properties in three dimensions.}
	{\AUT{A. Robledo, C. Varea}  
 	Proceedings of the CAM-94 Physics Meeting,
       	\emph{AIP Conf. Proc.} \textbf{342}, pp. 743-749, 1995.}
  \or
    \clasif{\bf ma 95.}
        {Can the Helfrich free energy for curved interfaces be derived from first principles?.}
	{\AUT{A. Robledo, C. Varea}  
       	  \emph{Physica A} \textbf{231}, 178-190 (1996).}
        \url{https://doi.org/10.1016/0378-4371(95)00457-2}
  \or
    \clasif{\bf aa 96.}
        {Stress tensor of inhomogeneous fluids.}
	{\AUT{C. Varea, A. Robledo}  
          \emph{Physica A} \textbf{233}, 132-144 (1996).}
        \url{https://doi.org/10.1016/S0378-4371(96)00244-0}
  \or      
    \clasif{\bf ma 97.}
        {Interfacial and capillary properties of a micellar solution model.}
	{\AUT{C. Varea, C. Garc\'{\i}a-Alc\'antara, A. Robledo}  
          \emph{Physica A} \textbf{236}, 177-187 (1997).}
        \url{https://doi.org/10.1016/S0378-4371(96)00397-4}
  \or
    \clasif{\bf ma 98.}
        {Scaling of interfacial tension and identity of bending moduli of microemulsions.}
	{\AUT{A. Robledo, C. Varea}  
        International School of Physics “Enrico Fermi”, Course CXXXIV, “The
	Physics of Complex Fluids”, julio 9-19 1996, Varenna, Italia, IOS Press 
        1997, pp 417-431}
  \or
    \clasif{\bf aa 99.}
        {Spinodal decomposition under confinement.}
	{\AUT{C. Varea, J. Campos-Ter\'an, A. Robledo}  
          \emph{Physica A} \textbf{244}, 440-452 (1997).}
        \url{https://doi.org/10.1016/S0378-4371(97)00228-8}
  \or
    \clasif{\bf aa 100.}
        {Interfacial width and shape fluctuations and extensions of the gaussian model of capillary waves.}
	{\AUT{A. Robledo, C. Varea}  
   	  \emph{J. Stat. Phys. } \textbf{89}, 273-282 (1997).}
        \url{https://doi.org/10.1007/BF02770765}
  \or
    \clasif{\bf aa 101.}
        {Phase properties of nematics confined by competing walls.}
	{\AUT{J. Quintana, A. Robledo}  
          \emph{Physica A} \textbf{248}, 28-43 (1998).}
        \url{https://doi.org/10.1016/S0378-4371(97)00523-2}
  \or
    \clasif{\bf aa 102.}
        {Fluctuations and instabilities of curved interfaces.}
	{\AUT{C. Varea, A. Robledo}  
          \emph{Physica A} \textbf{255}, 269-284 (1998)}
        \url{https://www.academia.edu/52441404}
  \or
    \clasif{\bf ma 103.}
        {Arrested phase separation and phase equilibrium properties of microemulsions.}
	{\AUT{A. Robledo, C. Varea}  
	  “International Workshop on the Morphology and Kinetics of Phase Separating Complex Fluids”,
          Messina, Italia, junio 24-28 1997. Il Nuovo Cimento 20 D, 2315-2324 (1998)}
  \or
    \clasif{\bf ma 104.}
        {Enhanced amphiphile adsorption at liquid-liquid interfaces in a micellar solution model.}
	{\AUT{C. Garc\'{\i}a-Alc\'antara, C. Varea, A. Robledo}  
          \emph{Physica A} \textbf{256}, 321-332 (1998)}
        \url{https://doi.org/10.1016/S0378-4371(98)00196-4}
  \or
    \clasif{\bf aa 105.}
        {Landau density functional theory for one-dimensional inhomogeneities.}
	{\AUT{A. Robledo, J. Quintana}  
          \emph{Physica A} \textbf{257}, 197-206 (1998)}
        \url{https://doi.org/10.1016/S0378-4371(98)00140-X}
  \or
    \clasif{\bf aa 106.}
        {Kinetics of phase change in binary mixtures with complete wetting interfaces.}
	{\AUT{J. Quintana, A. Robledo}  
          \emph{Mol. Phys.} \textbf{95}, 587-593 (1998)}
        \url{https://doi.org/10.1080/00268979809483192}
  \or
    \clasif{\bf aa 107.}
        {Density fluctuations and correlations of confined fluids.}
        {\AUT{C. Varea, A. Robledo}  
          \emph{Physica A} \textbf{268}, 391-411 (1999)}
        \url{https://doi.org/10.1016/S0378-4371(99)00049-7}
  \or
    \clasif{\bf d 108.}
       {?`Que pasa cuando se confina un fluido en un capilar? “Una ventana hacia la Investigaci\'on en F\'{\i}sica”.}
        {\AUT{A. Robledo}  
        Seminarios del Departamento de F\'{\i}sica Experimental en memoria de Angel Dacal Alonso,
        IFUNAM, mayo 22 1997, Compilador Dr. Esbaide Adem,
        Fondo de Cultura Econ\'omica (1999) pp252-257}
  \or
    \clasif{\bf aa 109.}
        {Renormalization group, entropy optimization, nonextensivity at criticality.}
        {\AUT{A. Robledo}  
        \emph{Phys. Rev. Lett.} \textbf{83}, 2289-2292 (1999)
        \url{https://doi.org/10.1103/PhysRevLett.83.2289}}
  \or
    \clasif{\bf aa 110.}
        {The renormalization group and optimization of entropy.}
        {\AUT{A. Robledo}  
          \emph{J. Stat. Phys.} \textbf{100}, 475 (2000).}
        \url{https://doi.org/10.1023/A:1018620618862}
  \or
    \clasif{\bf aa 111.}
        {Fluctuations and instabilities of model amphiphile interfaces.}
        {\AUT{C. Varea, A. Robledo}  
        \emph{Physica A} \textbf{290}, 360-378 (2001).}
        \url{https://doi.org/10.1016/S0378-4371(00)00461-1}
  \or
    \clasif{\bf aa 112.}
        {Anomalous diffusion, the renormalization group and optimization of entropy.}
        {\AUT{A. Robledo, J. Quintana}  
          \emph{Granular Matter} \textbf{3}, 29-32 (2001).}
        \url{https://doi.org/10.1007/s100350000064}
  \or
    \clasif{\bf aa 113.}
        {Confinement induced immiscibility of mixtures of enantiomers.}
        {\AUT{J. Quintana, A. Robledo}  
          \emph{Physica A} \textbf{295}, 333-347 (2001).}
        \url{https://doi.org/10.1016/S0378-4371(01)00129-7}
  \or
    \clasif{\bf aa 114.}
        {Theory of interfacial bending constants.}
        {\AUT{C. Varea, A. Robledo}  
          \emph{J. Phys.: Cond. Matt.} \textbf{13}, 9075-9088 (2001).}
        \url{https://doi.org/10.1088/0953-8984/13/41/303}
  \or
    \clasif{\bf aa 115.}
        {Scale-invariant random walks and optimization of non-extensive entropy.}
        {\AUT{A. Robledo, J. Quintana}  
          \emph{Chaos, Solitons and Fractals} \textbf{13}, 521-528 (2002)}
        \url{https://doi.org/10.1016/S0960-0779(01)00035-2}
  \or
    \clasif{\bf ma 116.}
        {“Phase transitions induced or suppressed by confinement”.}
        {\AUT{A. Robledo, J. Quintana}  
        NATO Advanced Research Series: ``New kinds of phase transitions: transformations in disordered substances'',
        Kluwer Academic Publishers 2002, pp. 545-555.}
  \or
    \clasif{\bf aa 117.}
        {“Stability of curved amphiphilic interfaces”.}
        {\AUT{C. Varea, A. Robledo}  
          \emph{Physica A} \textbf{306}, 301-315 (2002).}
        \url{https://doi.org/10.1016/S0378-4371(02)00507-1}
  \or
    \clasif{\bf aa 118.}
        {“Surface transitions under confinement”.}
        {\AUT{J. Quintana, A. Robledo}  
          \emph{J. Phys.: Cond. Matt.} \textbf{14}, 2211-2221 (2002).}
        \url{https://doi.org/10.1088/0953-8984/14/9/310}
  \or
    \clasif{\bf aa 119.}
        {“Universal renormalization-group dynamics at the onset of chaos in logistic maps and non-extensive statistical mechanics”.}
        {\AUT{F. Baldovin, A. Robledo}  
        \emph{Phys. Rev. E} \textbf{66}, 045104-1-4 (R) (2002).
          \url{https://www.doi.org/10.1103/PhysRevE.66.045104}}
  \or
    \clasif{\bf aa 120.}
        {“Sensitivity to initial conditions at bifurcations in one-dimensional non-linear maps: rigorous non-extensive solutions”.}
        {\AUT{F. Baldovin, A. Robledo}  
        \emph{EPL} \textbf{60}, 518-524 (2002).
          \url{https://doi.org/10.1209/epl/i2002-00249-7}}
  \or
    \clasif{\bf aa 121.}
        {“The renormalization group and optimization of non-extensive entropy: criticality in non-linear one-dimensional maps”.}
        {\AUT{A. Robledo}  
        \emph{Physica A} \textbf{314}, 437-441 (2002).}
  \or
    \clasif{\bf aa 122.}
        {“Revisiting disorder and Tsallis statistics”.}
        {\AUT{V. Latora, A. Rapisarda, A. Robledo}  
        \emph{Letters to the Editor, Science} \textbf{300}, 250 (2003).
        \url{https://doi.org/10.1126/science.300.5617.249d}}
  \or
    \clasif{\bf ma 123.}
        {“Unifying laws in multi-disciplinary power-law phenomena: fixed-point universality and non-extensive entropy”.}
        {\AUT{A. Robledo}  
   in: Non-extensive Entropy-Interdisciplinary Applications,
   Oxford University Press, C. Tsallis and M. Gell-Mann, editors (2004) pp. 63-78.}
        \url{https://doi.org/10.1093/oso/9780195159769.003.0008}
  \or
    \clasif{\bf aa 124.}
        {“Criticality in non-linear one-dimensional maps: RG universal map and non-extensive entropy”.}
        {\AUT{A. Robledo}  
        \emph{Physica D} \textbf{193}, 153-160 (2004).
          \url{https://doi.org/10.1016/j.physd.2004.01.016}}
  \or
    \clasif{\bf aa 125.}
        {“Nonextensive Pesin identity. Exact renormalization group analytical results for the dynamics at the edge of chaos of the logistic map”.}
        {\AUT{F. Baldovin, A. Robledo}  
        \emph{Phys. Rev. E} \textbf{69}, 045202(R) 1-4 (2004).
        \url{https://www.doi.org/10.1103/PhysRevE.69.045202}}
  \or
    \clasif{\bf ma 126.}
        {“Ubiquity of metastable-to-stable crossover in weakly chaotic dynamical systems”.}
        {\AUT{F. Baldovin, L.G. Moyano, A.P. Majtey, A. Robledo, C. Tsallis}  
        \emph{Physica A} \textbf{340}, 205-218 (2004).}
  \or
    \clasif{\bf ma 127.}
        {“Multifractality and nonextensivity at the edge of chaos of unimodal maps”.}
        {\AUT{E. Mayoral, A. Robledo}  
        \emph{Physica A} \textbf{340}, 219-226 (2004).}
  \or
    \clasif{\bf aa 128.}
        {“Universal glassy dynamics at noise-perturbed onset of chaos. A route to ergodicity breakdown”.}
        {\AUT{A. Robledo}  
        \emph{Phys. Lett. A} \textbf{328}, 467-472 (2004).
        \url{https://doi.org/10.1016/j.physleta.2004.06.062}}
  \or
    \clasif{\bf ma 129.}
        {“Aging at the edge of chaos: Glassy dynamics and nonextensive statistics”.}
        {\AUT{A. Robledo}  
        \emph{Physica A} \textbf{342}, 104-111 (2004).
        \url{https://doi.org/10.1016/j.physa.2004.04.065}}
  \or
    \clasif{\bf ma 130.}
        {“Critical fluctuations, intermittent dynamics and Tsallis statistics”.}
        {\AUT{A. Robledo}  
          \emph{Physica A} \textbf{344}, 631-636 (2004).}
        \url{https://doi.org/10.1016/j.physa.2004.06.043}
  \or
    \clasif{\bf ma 131.}
        {“Intermittency at critical transitions and aging dynamics at edge of chaos“.}
        {\AUT{A. Robledo}  
        \emph{Pramana} \textbf{64}, 947-956 (2005).
        \url{https://www.ias.ac.in/article/fulltext/pram/064/06/0947-0956}}
  \or
    \clasif{\bf ma 132.}
        {“Glassy dynamics at the onset of chaos with additive noise“.}
        {\AUT{F. Baldovin, A. Robledo}  
        \emph{Fluctuation and Noise Letters} \textbf{5}, L313-L318 (2005).
        \url{https://doi.org/10.1142/S0219477505002707}}
  \or
    \clasif{\bf ma 133.}
        {“Glassy dynamics at the edge of chaos”.}
	{\AUT{A. Robledo}  
        in “Current Topics in Physics”, 
        Imperial College Press (ICP), R. Barrio and K. Kaski, editores. Cap\'{\i}tulo 5, pp 83-93, 2005.}
  \or
    \clasif{\bf aa 134.}
        {“Unorthodox properties of critical clusters”.}
	{\AUT{A. Robledo}  
        \emph{Mol. Phys.} \textbf{103}, 3025 (2005).
        \url{https://doi.org/10.1080/00268970500185989}}
  \or
    \clasif{\bf aa 135.}
        {“Tsallis' $\bs{q}$ index and Mori's $\bs{q}$ phase transitions at edge of chaos“.}
	{\AUT{E. Mayoral, A. Robledo}  
        \emph{Phys. Rev. E} \textbf{72}, 026029 (2005).
        \url{https://www.doi.org/10.1103/PhysRevE.72.026209}}
  \or
    \clasif{\bf aa 136.}
        {“Parallels between the dynamics at the noise-perturbed onset of chaos in logistic maps and the dynamics of glass formation”.}
        {\AUT{F. Baldovin, A. Robledo}  
        \emph{Phys. Rev. E} \textbf{72}, 066213 (2005).
        \url{https://www.doi.org/10.1103/PhysRevE.72.066213}}
  \or
    \clasif{\bf aa 137.}
        {“Critical attractors and $\bs{q}$-statistics”.}
	{\AUT{A. Robledo}  
        in ``Nonextensive Statistical Mechanics: New Trends, New Perspectives'', 	JP Boon and C. Tsallis, editores
        \emph{Europhys. News} \textbf{36}, 214-218 (2005).
        \url{https://doi.org/10.1051/epn:2005611}}
  \or
    \clasif{\bf ma 138.}
        {“Two stories outside Boltzmann-Gibbs statistics: Mori's $\bs{q}$-phase transitions and glassy dynamics at the onset of chaos“.}
        {\AUT{A. Robledo, F. Baldovin, E. Mayoral}  
        in Complexity, Metastability and Nonextensivity, 
        World Scientific, 2005, 43-54,
        \url{https://doi.org/10.1142/9789812701558_0004}}
  \or
    \clasif{\bf ma 139.}
        {“A recent appreciation of the singular dynamics at the edge of chaos“.}
        {\AUT{E. Mayoral, A. Robledo}  
        in Verhulst 200 on Chaos, Understanding Complex Systems (Series), M. Ausloss and M. Dirickx, editores.
        (Springer, Berlin, 2006) pp. 339-354.}
  \or
    \clasif{\bf ma 140.}
        {“Crossover from critical to chaotic attractor dynamics in logistic and circle maps”.}
        {\AUT{A. Robledo}  
          Prog. Theor. Phys. Supp. 162, 10-17 (2006).
        \url{https://doi.org/10.1143/PTPS.162.10}}
  \or
    \clasif{\bf ma 141.}
        {“Critical attractors and the physical realm of $\bs{q}$-statistics”.}
        {\AUT{A. Robledo}  
        in ``Chaos, Nonlinearity, Complexity: The dynamical paradigm of nature''
           A. Sengupta, editor, Springer-Verlag, 2006, pp72-113.}
  \or
    \clasif{\bf d 142.}
       {“G\'enesis de una nueva f\'{\i}sica estad\'{\i}stica”.}
        {\AUT{A. Robledo}  
        in “Descubrimientos y Aportaciones Cient\'{\i}ficas y Human\'{\i}sticas Mexicanas en el Siglo Veinte”, Sergio Estrada Orihuela, editor,
           Academia Mexicana de Ciencias/ Fondo de Cultura Econ\'omica, 2006.}
  \or
    \clasif{\bf aa 143.}
        {“Dynamics at the quasiperiodic onset of chaos, Tsallis $\bs{q}$-statistics and Mori's $\bs{q}$-phase thermodynamics”.}
        {\AUT{H. Hern\'andez-Salda\~na, A. Robledo}  
        \emph{Physica A} \textbf{370}, 286-300 (2006).
          \url{https://doi.org/10.1016/j.physa.2006.03.018}}
  \or
    \clasif{\bf aa 144.}
        {“Incidence of nonextensive thermodynamics in temporal scaling at Feigenbaum points and non-extensive thermodynamics”.}
        {\AUT{A. Robledo}  
        \emph{Physica A} \textbf{370}, 449-460 (2006).
        \url{https://doi.org/10.1016/j.physa.2006.06.003}}
  \or
    \clasif{\bf ma 145.}
        {“$\bs{q}$-statistical properties of large critical clusters”.}
        {\AUT{A. Robledo}  
        \emph{Int. J. Mod. Phys. B} \textbf{21}, 3947-3953 (2007).
        \url{https://doi.org/10.1142/S0217979207045001}}
  \or
    \clasif{\bf ma 146.}
        {“Some aspects of the dynamics towards supercycle attractors and their accumulation point, the Feigenbaum attractor“.}
        {\AUT{A. Robledo, L.G. Moyano}  
        \emph{AIP Con. Proc.} \textbf{965}, (2007) 114
        \url{https://doi.org/10.1063/1.2828722}}
  \or
    \clasif{\bf aa 147.}
        {“$\bs{q}$-deformed statistical-mechanical property in the dynamics of trajectories en route to the Feigenbaum attractor”.}
	{\AUT{A. Robledo, L.G. Moyano}  
        \emph{Phys. Rev. E} \textbf{77}, 032613-1-14 (2008).
        \url{https://doi.org/10.1103/PhysRevE.77.036213}}
  \or
    \clasif{\bf ma 148.}
        {“Labyrinthine pathways towards supercycle attractors in unimodal maps”.}
	{\AUT{L.G. Moyano, D. Silva, A. Robledo}  
        \emph{Cent. Eur. J. Phys.} \textbf{7}, 591-600 (2009).
        \url{https://doi.org/10.2478/s11534-009-0065-1}}
  \or
    \clasif{\bf ma 149.}
        {“Dynamics towards the Feigenbaum attractor”.}
	{\AUT{A. Robledo, L.G. Moyano}  
        \emph{Braz. J. Phys.} \textbf{39}, 364-370 (2009).
        \url{https://doi.org/10.1590/S0103-97332009000400004}}
  \or
    \clasif{\bf ma 150.}
        {“Generalized thermodynamics underlying the laws of Zipf and Benford”.}
	{\AUT{C. Altamirano, A. Robledo}  
         (Springer-Verlag, LNICST 5 2009) pp. 2232-2237,
        \url{https://doi.org/10.1007/978-3-642-02469-6_100}}
  \or
    \clasif{\bf aa 151.}
        {“Equivalence between the mobility edge of electronic transport on disorderless networks and the onset of chaos via intermittency in deterministic maps”.}
        {\AUT{M. Mart\'{i}nez-Mares, A. Robledo}  
    	\emph{Phys. Rev. E} \textbf{80}, 045201(R)-1-4 (2009).
        \url{https://www.doi.org/10.1103/PhysRevE.80.045201}}
  \or
    \clasif{\bf aa 152.}
        {“Renormalization Group structure for sums of variables generated by incipiently chaotic maps”.}
	{\AUT{M.A. Fuentes, A. Robledo}  
        \emph{J. Stat. Mech. Theory Exp.} \textbf{2010}, P01001 (2010).
        \url{https://doi.org/10.1088/1742-5468/2010/01/P01001}}
  \or
    \clasif{\bf ma 153.}
        {“$\bs{q}$-deformed statistical-mechanical structure in the dynamics of the Feigenbaum attractor”.}
        {\AUT{A. Robledo}  
        \emph{J. Phys. Conf. Ser.} \textbf{246}, 012025+6 9 (2010).
        \url{https://doi.org/10.1088/1742-6596/246/1/012025}}
  \or
    \clasif{\bf ma 154.}
        {Stationary distributions of sums of marginally chaotic variables as renormalization group fixed points”.}
        {\AUT{M.A. Fuentes, A. Robledo}  
        \emph{J. Phys.: Conf. Ser.} \textbf{201}, 012002 (2010).
        \url{https://doi.org/10.1088/1742-6596/201/1/012002}}
  \or
    \clasif{\bf aa 155.}
        {“Renewal stochastic processes with correlated events. Phase transitions along time evolution”.}
        {\AUT{J. Vel\'azquez, A. Robledo}  
        \emph{Phys. Rev. E} \textbf{83}, 031103-1-031103-6 (2011).}
        \url{https://doi.org/10.1103/PhysRevE.83.031103}
  \or
    \clasif{\bf aa 156.}
        {“Possible thermodynamic structure underlying the laws of Zipf and Benford”.}
        {\AUT{C. Altamirano, A. Robledo}  
        \emph{Eur. Phys. J. B} \textbf{81}, 345-351 (2011).
        \url{https://doi.org/10.1140/epjb/e2011-10968-5}}
  \or
    \clasif{\bf ma 157.}
        {“Laws of Zipf and Benford, intermittency and critical fluctuations”.}
        {\AUT{A. Robledo}  
        \emph{ Chin. Sci. Bull.} \textbf{56}, 3643-3648 (2011).
        \url{https://doi.org/10.1007/s11434-011-4827-y}}
  \or
    \clasif{\bf aa 158.}
        {“Feigenbaum graphs: a complex network perspective of chaos”.}
        {\AUT{B. Luque, L. Lacasa, F.J. Ballesteros, A. Robledo}  
          \emph{PLoS ONE} \textbf{6}(9): e22411 (2011).
        \url{https://doi.org/10.1371/journal.pone.0022411}}
  \or
    \clasif{\bf aa 159.}
        {Chaos and unpredictability in evolutionary dynamics in discrete time.}
        {\AUT{D. Vilone, A. Robledo, A. S\'anchez}  
        \emph{Phys. Rev. Lett.} \textbf{107}, 038101-1-038101-4 (2011).
        \url{https://doi.org/10.1103/PhysRevLett.107.038101}}
  \or
    \clasif{\bf aa 160.}
        {“Statistical-mechanical structure for renewal stochastic processes”.}
        {\AUT{J. Vel\'azquez, A. Robledo}  
        \emph{IJAMAS} \textbf{26}, 3-15 (2012).}
        \url{https://www.researchgate.net/publication/215439762}
  \or
    \clasif{\bf aa 161.}
        {“Analytical properties of horizontal visibility graphs in the Feigenbaum scenario”.}
        {\AUT{B. Luque, L. Lacasa, F.J. Ballesteros, A. Robledo}  
        \emph{Chaos} \textbf{22}, 013109-013109-14 (2012).
        \url{https://doi.org/10.1063/1.3676686}}
  \or
    \clasif{\bf aa 162.}
        {“M\"obius transformations and electronic transport properties of large networks”.}
        {\AUT{Yu Jiang, M. Mart\'{\i}nez-Mares, E. Casta\~no, A. Robledo}  
        \emph{Phys. Rev. E} \textbf{85}, 057202-057202-4 (2012).
        \url{https://doi.org/10.1103/PhysRevE.85.057202}}
  \or
    \clasif{\bf aa 163.}
        {“Feigenbaum graphs at the onset of chaos”.}
        {\AUT{B. Luque, L. Lacasa, A. Robledo}  
        \emph{Phys. Lett. A} \textbf{376}, 3625--3629 (2012).
        \url{https://doi.org/10.1016/j.physleta.2012.10.050}}
  \or
    \clasif{\bf aa 164.}
        {“A dynamical model for hierarchy and modular organization: The trajectories en route to the attractor at the transition to chaos”.}
        {\AUT{A. Robledo}  
        \emph{J. Phys. Conf. Ser.} \textbf{394}, 012007-012007-9 (2012).
        \url{https://doi.org/10.1088/1742-6596/394/1/012007}}
  \or
    \clasif{\bf aa 165.}
        {Quasiperiodic graphs: structural design, scaling and entropic properties.}
        {\AUT{B. Luque, A. N\'u\~nez, F.J. Ballesteros, A. Robledo}  
        \emph{J. Nonlinear} \textbf{Sci.}, 23, 335-342 (2013).
        \url{https://doi.org/10.1007/s00332-012-9153-2}}
  \or
    \clasif{\bf aa 166.}
        {Horizontal visibility graphs generated by type-I intermittency.}
        {\AUT{A. N\'u\~nez, B. Luque, L. Lacasa, J.P. G\'omez, A. Robledo}  
        \emph{Phys. Rev. E} \textbf{87}, 052801-052801-9 (2013).
        \url{https://www.doi.org/10.1103/PhysRevE.87.052801}}
  \or
    \clasif{\bf aa 167.}
        {Quasiperiodic graphs at the onset of chaos.}
        {\AUT{A. N\'u\~nez, B. Luque, M. Cordero, M. G\'omez, A. Robledo}  
        \emph{Phys. Rev. E} \textbf{88}, 06918-1-06918-8 (2013).
        \url{https://www.doi.org/10.1103/PhysRevE.88.062918}}
  \or
    \clasif{\bf aa 168.}
        {Generalized Statistical Mechanics at the Onset of Chaos.}
        {\AUT{A. Robledo}  
        \emph{Entropy} \textbf{15}, 5178-5222 (2013)
          \url{https://doi:10.3390/e15125178}}
  \or
    \clasif{\bf aa 169.}
        {Sums of variables at the onset of chaos.}
        {\AUT{M.A. Fuentes, A. Robledo}  
        \emph{Eur. Phys. J. B} \textbf{87}, 32 (2014) \url{https://doi:10.1140/epjb/e2014-40882-1}}
  \or
    \clasif{\bf aa 170.}
        {Emergent statistical-mechanical structure in the dynamics along the period-doubling route to chaos.}
        {\AUT{A. Diaz-Ruelas, A. Robledo}  
        \emph{EPL} \textbf{105}, 40004 (2014).
        \url{https://doi.org/10.1209/0295-5075/105/40004}}
  \or
    \clasif{\bf aa 171.}
        {Scaling of distributions of sums of positions for chaotic dynamics at band-splitting points.}
        {\AUT{A. Diaz-Ruelas, M.A. Fuentes, A. Robledo}  
        \emph{EPL} \textbf{108}, 20008 (2014).
          \url{https://doi.org/10.1209/0295-5075/108/20008}}
  \or
    \clasif{\bf aa 172.}
        {Incidence of $\bs{q}$ statistics in rank distributions.}
        {\AUT{C. Yalcin, A. Robledo, M. Gell-Mann}  
        \emph{PNAS} \textbf{111}, (2014).
          \url{https://doi.org/10.1073/pnas.1412093111}}
  \or
    \clasif{\bf ma 173.}
        {Pascal (Yang Hui) triangles and power laws in the logistic map.}
	{\AUT{C. Velarde, A. Robledo}  
        \emph{J. Phys. Conf. Ser.} \textbf{604}, 012018-012018-7 (2015).
        \url{https://doi.org/10.1088/1742-6596/604/1/012018}}
  \or
    \clasif{\bf aa 174.}
        {Entropies for severely contracted configuration space.}
        {\AUT{G.C. Yalcin, C. Velarde, A. Robledo}  
        \emph{Heliyon} \textbf{1}, (2015) e00045,
          \url{https://doi.org/10.1016/j.heliyon.2015.e00045}}
  \or
    \clasif{\bf aa 175.}
        {Entropy and renormalization in chaotic visibility graphs, en Mathematical Foundations and Applications of Graph Entropy.}
        {\AUT{B. Luque, F.J. Ballesteros, A. Robledo, L. Lacasa}  
        \emph{Wiley Online Library}, \textbf{6}, (2017) 1-39,
        \url{https://doi.org/10.1002/9783527693245.ch1}}
  \or
    \clasif{\bf ma 176.}
        {Sums of variables at the onset of chaos, replenished.}
        {\AUT{A. Diaz-Ruelas, A. Robledo}  
        \emph{EPJ-ST Special Issue 160011: Temporal and Spatio-Temporal Dynamic Instabilities: Novel Computational and Experimental Approaches
        Eur. Phys. J. Spec. Top.} \textbf{225}, 2763 (2016).
        \url{https://doi.org/10.1140/epjst/e2016-60011-y}}
  \or
    \clasif{\bf ma 177.}
        {Tangent map intermittency as an approximate analysis of intermittency in a high dimensional fully stochastic dynamical system: The Tangled Nature model.}
        {\AUT{A. Diaz-Ruelas, A. Robledo, H.J. Jensen, D. Piovani}  
        \emph{Chaos} \textbf{26}, 123105 (2016)
        \url{https://doi.org/10.1063/1.4968207}}
  \or
    \clasif{\bf ma 178.}
        {Typical length scales in conducting disorderless networks.}
        {\AUT{M. Mart\'{i}nez-Mares, V. Dominguez-Rocha, A. Robledo}  
        EPJ-ST/172 Special Issue: Nonlinearity, Nonequilibrium and Complexity: Questions and Perspectives in Statistical Physics
        \emph{Eur. Phys. J. Spec. Top.} \textbf{226} 417 (2017)
        \url{https://doi.org/10.1140/epjst/e2016-60129-x}}
  \or
    \clasif{\bf ma 179.}
        {Dual characterization of critical fluctuations: Density functional theory \& nonlinear dynamics close to a tangent bifurcation.}
        {\AUT{M. Riquelme-Galvan, A. Robledo}  
        EPJ-ST/172 Special Issue: Nonlinearity, Nonequilibrium and Complexity: Questions and Perspectives in Statistical Physics
        \emph{Eur. Phys. J. Spec. Top.} \textbf{226} 433 (2017)
        \url{https://doi.org/10.1140/epjst/e2016-60268-0}}
  \or
    \clasif{\bf ma 180.}
        {Relating high dimensional stochastic complex systems to low-dimensional intermittency?.}
         {\AUT{A. Diaz-Ruelas, A. Robledo, H.J. Jensen, D. Piovani}  
        EPJ-ST/172 Special Issue: Nonlinearity, Nonequilibrium and Complexity: Questions and Perspectives in Statistical Physics
        \emph{Eur. Phys. J. Spec. Top.} 226 341 (2017)
        \url{https://doi.org/10.1140/epjst/e2016-60264-4}}
  \or
    \clasif{\bf aa 181.}
        {Rank distributions: Frequency vs. Magnitude.}
        {\AUT{C. Velarde, A. Robledo}  
        \emph{PLoS ONE} 12(10): e0186015 (2017). \url{https://doi.org/10.1371/journal.pone.0186015}}
  \or
    \clasif{\bf aa 182.}
        {Manifestations of the onset of chaos in condensed matter and complex systems.}
        {\AUT{C. Velarde, A. Robledo}  
        \emph{Eur. Phys. J. Spec. Top.} 227 645 (2018).
        \url{https://doi.org/10.1140/epjst/e2018-00128-9}}
  \or
    \clasif{\bf aa 183.}
        {Invariant density of intermittent nonlinear maps descriptive of coherent quantum transport through disorderless lattices.}
        {\AUT{V. Dominguez-Rocha, R.A. Mendez-Sanchez, M. Mart\'{i}nez-Mares, A. Robledo}  
        \emph{Physica D} \textbf{412}, (2020) 132623, \url{https://doi.org/10.1016/j.physd.2020.132623}}
  \or
    \clasif{\bf aa 184.}
        {Dynamical analogues of rank distributions.}
        {\AUT{C. Velarde, A. Robledo}  
        PLoS ONE 14(2): e0211226 (2019).
        \url{https://doi.org/10.1371/journal.pone.0211226}}
  \or
    \clasif{\bf aa 185.}
        {Self-organization and a constrained thermal system analogue of the onset of chaos.}
        {\AUT{A. Robledo}  
        \emph{EPL} \textbf{123}, 40003 (2018).
        \url{https://doi.org/10.1209\%2F0295-5075\%2F123\%2F40003}}
  \or
    \clasif{\bf aa 186.}
        {Statistical mechanical model for growth and spread of contagions under gauged population confinement.}
        {\AUT{C. Velarde, A. Robledo}  
        \emph{Physica A} \textbf{573}, 125960 (2021).
        \url{https://doi.org/10.1016/j.physa.2021.125960}}
  \or
  \clasif{\bf aa 187.}
         {Logistic map trajectory distributions: Renormalization-group, entropy,
             and criticality at the transition to chaos.}
           {\AUT{A. Diaz-Ruelas, F. Baldovin, A. Robledo}  
             \emph{Chaos}, \textbf{31}, 033112 (2021).
             \url{https://doi.org/10.1063/5.0040544}}
  \or
    \clasif{\bf aa 188.}
        {Number theory, borderline dimension and extensive entropy in distributions
          of ranked data.}
        {\AUT{C. Velarde, A. Robledo}  
          PLos ONE 17(12): e0279448 (2022).
        \url{https://doi.org/10.1371/journal.pone.0279448}}
  \or
    \clasif{\bf aa 189.}
       {Publication under elaboration.}
        {Link to seminar:\\
       \url{https://drive.google.com/file/d/1HaZzU8VnlKf5qO6miVWQ3Gc_pw5rqvdO/view}}
  \or
    \clasif{\bf aa 190.}
       {G\'enesis de una nueva f\'{i}sica estad\'{i}stica.}
       {\AUT{A. Robledo}  
        En Descubrimientos y Aportaciones Cient\'{i}ficas y
           Human\'{i}sticas Mexicanas en el Siglo Veinte,
           Octavio Paredes L\'opez, Sergio Estrada Orihuela, editores,
           Academia Mexicana de Ciencias/ Fondo de Cultura Econ\'omica,
           M\'exico, pp. 818-829 (2006). ISBN: 9789681686345}
  \or
    \clasif{\bf aa 191.}
     {How, Why and When Tsallis Statistical Mechanics Provides Precise Descriptions
       of Natural Phenomena.}
     {\AUT{A. Robledo, C. Velarde}  
     \emph{Entropy} \textbf{24}, 1761 (2022).
      \url{https://doi.org/10.3390/e24121761}}
  \else ?\fi} }

\def\nCls#1{
  \ifcase #1
         \or aa \or aa \or aa \or aa \or aa \or aa \or aa \or aa \or aa 
  \or aa \or aa \or aa \or aa \or aa \or aa \or ma \or m  \or aa \or aa
  \or m  \or m  \or m  \or e  \or e  \or d  \or aa \or aa \or aa \or aa
  \or m  \or aa \or aa \or aa \or aa \or ma \or e  \or aa \or aa \or aa
  \or ma \or aa \or m  \or aa \or ma \or ma \or aa \or ma \or aa \or aa 
  \or e  \or e  \or d  \or aa \or aa \or aa \or aa \or ma \or ma \or ma
  \or ma \or m  \or m  \or aa \or aa \or ma \or m  \or m  \or e  \or d
  \or aa \or aa \or aa \or ma \or aa \or m  \or m  \or aa \or aa \or aa
  \or aa \or m  \or ma \or aa \or aa \or aa \or aa \or aa \or aa \or ma
  \or m  \or aa \or aa \or ma \or ma \or ma \or aa \or ma \or ma \or aa 
  \or aa \or aa \or aa \or ma \or ma \or aa \or aa \or aa \or d  \or aa
  \or aa \or aa \or aa \or aa \or aa \or aa \or ma \or aa \or aa \or aa
  \or aa \or aa \or aa \or ma \or aa \or aa \or ma \or ma \or aa \or ma
  \or ma \or ma \or ma \or ma \or aa \or aa \or aa \or aa \or ma \or ma
  \or ma \or ma \or d  \or aa \or aa \or ma \or ma \or aa \or ma \or ma 
  \or ma \or aa \or aa \or ma \or ma \or aa \or aa \or ma \or aa \or aa
  \or aa \or aa \or aa \or aa \or aa \or aa \or aa \or aa \or aa \or aa
  \or aa \or aa \or aa \or ma \or aa \or aa \or ma \or ma \or ma \or ma
  \or ma \or aa \or aa \or aa \or aa \or aa \or aa \or aa \or aa \or aa
  \or aa \or aa                                                          
  \else ?\fi}                                                                                                                                  

\def\clasif#1{}  
\def\AUT#1{#1.}   

\def\Abs#1{\preparaColorPub{#1}                                      
  \subsubsection*{\refstepcounter{subsubsection}\label{\mysss}
                  \ColorPub\thesubsubsection\nPub(#1)}
}

\def\preparaColorPub#1{\StrLen{\nCls{#1}}[\MyStrLen]}
\def\ColorPub{\ifthenelse{\equal{\MyStrLen}{4}}{\color{blue}}{\color{blue}} }   

\def\NewT{}  \def\RED{}  \def\BLACK{}   \def\NEW#1{#1}

\begin{document}

\title{A Half-Century Research Footpath in Statistical Physics}

\author{Alberto Robledo\textsuperscript{1}, Carlos Velarde\textsuperscript{2}\\
  \footnotesize 1. Instituto de F\'isica, Universidad Nacional Aut\'onoma de M\'exico\\
  \footnotesize 2. Instituto de Investigaciones en Matem\'aticas Aplicadas y en Sistemas,\\
       \footnotesize           Universidad Nacional Aut\'onoma de M\'exico\\
}
  
\vskip-3\baselineskip   
\maketitle

\begin{abstract}  
  \baselineskip=0.93\baselineskip   

  We give an abridged account\remove[r1]{, assembling references to a selected list of publications,}
  of a continued string of studies in condensed
matter physics and in complex systems that span five decades.
\add[r1]{We provide links to access abstracts and full texts of a selected list of publications.}
The studies were carried out within a framework of methods and
models, some developed in situ, of \add[CV]{stochastic processes,} statistical \change[CV]{physics}{mechanics} and nonlinear
dynamics. The topics, techniques and outcomes reflect evolving
interests of the community but also show a particular character that
privileges the use of analogies or unusual viewpoints that unite the
studies in distinctive ways. We place the studies into three
collections according to the main underlying approach: stochastic
processes, density functional theory, and nonlinear dynamics. In the
first group we include the following: i) Random walks for fluid correlations.
ii) Random walks for electronic band structures. iii) Trapping of multiple
walkers. iv) Renormalization group and entropy for Weierstrass walks. v)
Statistical-mechanical analogy for renewal processes. vi) Phase transitions
along time in correlated renewal processes. In the second group: i) Density
functionals and Widom's particle insertion method. ii) Liquid to solid
transitions \change[CV]{of}{in} hard-core model systems. iii) Nucleation and spinodal
decomposition. iv) Global wetting phase diagrams. v) Lattice models
for complex fluids. vi) Anomalous micellar solution criticality. vii) Complex
fluids under confinement. viii) Density functionals for curved interfaces.
ix) Curvature interfacial transitions. x) Line tension and wetting. xi)
Classical and quantum mechanical analogs built in density functional
theory. xii) Phase behavior and pairing mechanism for two-dimensional
superconductors. In the third group: i) Critical fluctuations and the
route out of chaos. ii) Glassy dynamics at the noise-perturbed onset
of chaos. iii) Localization transition as a bifurcation. iv) Pesin identity
at the Feigenbaum attractor. v) Renormalization group for chaotic
attractors. vi) Self-organization along the
period-doubling cascade. vii) Universality classes of rank distributions.
viii) Complex network view of the routes to chaos. ix) Chaos in discrete
time game theory.
\add[r1]{x) Equivalence of paradigms (edge of chaos and criticality).
         xi)  Allometry in biology and human activities.}
xii) Nested systems. 
We discuss the body of knowledge created by these
research lines in relation to theoretical foundations and spread of
subjects. We indicate unsuspected connections underlying different
aspects of these investigations and also point out both natural and
unanticipated perspectives for future developments.
\add[r1]{Finally, we refer to our most important and recent contribution: An answer
with a firm basis to the long standing question about the limit of validity
of ordinary statistical mechanics and the pertinence of Tsallis statistics.}
\end{abstract}

\newpage

\tableofcontents   
\newpage

\section*{Introduction}
\addcontentsline{toc}{section}{Introduction}

Half a century ago the frontiers of statistical physics were about to
expand considerably and conquer new territories. Amongst them, and
most visibly, the understanding of the scaling laws of critical
phenomena exposed a few years earlier \cite{CriticalPhenomena1}. The
advancement was made possible via the development of the novel
research tool known as Renormalization Group
\cite{RenormalizationGroup1}. Concurrently, and with features parallel to
those of criticality, the nonlinear dynamical discovery of the routes
to deterministic chaos from regular \add[CV]{to irregular} behavior took place \cite{Schuster1}.
Alongside, powerful methods for stochastic processes were just
beginning to be applied to random walks and renewal processes
associated with long-tailed distributions as opposed to the customary
normal case \cite{Montroll4}. Equally, at that time new
statistical-mechanical techniques made possible a shift of focus from
the study of simple homogeneous states towards more complicated
situations. Contemporarily, studies of phase transitions in various
different settings such as those occurring within surfaces became
feasible, and the field of complex fluids, also referred to as soft
condensed matter, grew to be differentiated from statistical mechanics
of polymers \cite{ComplexFluids1}.  We shall not describe here, in general,
the evolution of statistical physics in the subsequent decades, but
information on this can be obtained from different sources, such as,
for example, the proceedings of the International Conferences on
Statistical Physics (Statphys) that have occurred with three-year
periodicity over this time \cite{StatphysConferences1}.

Five decades ago the Mexican research community in statistical physics
could be practically counted \change[CV]{by}{with} the fingers of one hand, but it was
already undergoing a strong process of growth, with new participants
naturally curious and positively influenced by the larger
international community interests. By the turn of the century, half
way through the 50-year period involved in this account, the Mexican
community in the field had increased and matured very
considerably. This was reflected by the organization and hosting of
the Statphys Conference in Cancun, Mexico, in 2001 \cite{Statphys21}, the
first \add[CV]{to be held} in a Spanish speaking country of these long series of flagship
conferences. In the present day, the Mexican community in the field
has developed much further and has become a large and vigorous
research group spread into an array of topics that represent the
current research pursuits in statistical physics and in frontier
interdisciplinary science.

Here we give an account of an uninterrupted trajectory of studies
based in Mexico and which covers this long period of development. The
studies started by linking stochastic processes, such as random walks,
to fluid or electronic structures, then \remove[r1]{considered} the trapping of
multiple walkers \add[r1]{was considered}, and much later,
\add[r1]{these studies} led to a Renormalization Group view
for long-tailed distributions\add[CV]{,} and to renewal processes undergoing                           
phase change along time. \change[r1]{O}{Ano}ther, larger set of studies, mainly based         
on the free energy density functional approach, \change[r1]{obtained}{produced} descriptions
for liquid-solid transitions, interfacial transitions, and kinetics of
phase change. \change[r1]{Then}{Subsequently}, models were constructed for micellar solutions,
microemulsions and magnetic alloys, the properties of various types of
confined systems, such as nematics and \add[r1]{mixtures of} quiral molecules\remove[r1]{solutions}, were
determined, and the understanding of magnetism and superconductivity
phase behavior in two-dimensional systems was advanced.
Latterly, \add[r1]{through an additional set of studies,} important features   
were uncovered on the dynamics at the onset of chaos in nonlinear
systems, which unexpectedly connected to difficult problems in
condensed matter physics, such as the dynamics close to glass
formation, localization phenomena and critical fluctuations. \change[r1]{Finally}{But also},
the same set of dynamical properties has been used for the design of
models for complex systems phenomena including self-organization.
\add[r1]{Finally, the accumulated experience gained through the studies
  involving density functional theory and iterated-time nonlinear low-dimensional
  dynamics led to a promising advance regarding an enduring question on a
  generalization of statistical mechanics, the so-called $q$-statistics.}

The studies are allocated into \change[r1]{twenty-seven}{thirtytwo} groups and these in turn
into \change[r1]{three}{four} main parts. Each part
\remove[CV]{(except the fourth)}  
has a preamble followed by its
sections (the groups of studies in the part). Each section is given an
appropriate informative title and it is followed by a numbered
collection of \add[r1]{links to abstracts and} references to selected advances that represent the
focused subject. \remove[r1]{We end our account with a summary of the most
relevant developments and an outlook of possible future work.} A
particular selected advance is located by the part (I, II or III), its
section (i, ii, iii, ...), and its explanatory comment and \change[r1]{reference}{links} (1,2,3,...).
\add[r1]{We end our account with a summary of the most
  relevant developments and an outlook of possible future work.}

\NewT{
A critical assessment is provided for each of the \change[r1]{27}{32} topics. The
assessments constitute the main informative content of \change[r1]{the}{this} review,
they indicate the basic advances accomplished but also their
limitations; the novelty or uniqueness of the studies, then too the
use of preceding knowledge and, or, previously developed methodologies
and models. The articles referenced in each assessment contain more
technical detail about these developments and provide specific
information already in the literature. The statistical physics
developments considered are in this case not all recent but extended
throughout several decades, not global but restricted to a specific
country, and not covering all topics but limited to a particular
experience.

\section{Stochastic Processes}\label{s-StochProc}  

\textbf{Random walk analogs.} Elliot Montroll advanced a particularly efficient
and elegant format to study random walks and renewal processes
\cite{Montroll1, Montroll2}. The linearity property of independent
identically distributed (iid) events \add[CV]{(on regular lattices when spatial positions are required)} allowed for a prominent role for
the generating functions of the basic sets of probabilities. Once determined,
Fourier, Laplace or $z$-transforms, \remove[CV]{or} their inverses, and use of
differentiation when necessary, facilitate\add[r1]{s} access to all conceivable
properties \cite{Montroll1, Montroll2}.  Recurrence relations are
expressible as convolutions. First passage times become trapping
events and traps can be arranged to form \add[CV]{lines,} surfaces, etc. Distinction
between time and number of events along a sequence of \change[CV]{them}{steps} leads to a
\change[r1]{seemly}{fitting} continuous time random \change[CV]{process}{walk} description.

How to make use of these stochastic processes in other fields? First,
we established analogies via the identification of convolutions
\change[CV]{amongst its}{present in the} basic expressions. One case is the Ornstein-Zernicke
relation, a common starting point for the development of integral
equation approa\add[CV]{c}hes for correlation functions descriptive of
equilibrium fluid structures. Another one is the Green function
relation for tight binding electronic band structures. Secondly, there
is the requirement for primary events in photosynthesis to be triggered,
\change[r1]{for}{only after} the trapping at a reaction site of more than one energy packet.
This led to the study of first passage times for multiple walkers.

The elegant Weierstrass walk, designed by Montroll and colleagues
\cite{Montroll3} to observe the emergence in the appropriate limits of
either the Gaussian or the Levy distributions, has a rich set of
scaling properties reminiscent of those of the nontrivial fixed points
of the Renormalization Group \add[r1]{(RG)} when applied to critical phenomena in
thermal systems. It was our interest to generalize this \change[CV]{single}{specific} walk
into an infinite family of them, that took the form of a universality class, to
see if it revealed more features of this likeness, such as the flow to
trivial fixed points. Moreover, it presented the opportunity of calculating
entropies for the entire family of step distributions, and observe if the
associated fixed points represent entropy extrema. In this way it was
possible to reveal a hidden connection between \change[r1]{this}{the RG} technique and
entropy optimization \cite{Montroll4}.

In the case of renewal processes it caught our attention the fact that
the expression for a key quantity resembled that of a grand canonical  
partition function in equilibrium statistical mechanics. This expression
is that for the Laplace transform of the generating function for the
probability density for the occurrence of $n$ events at time $t$.
This identification would imply that the probability density itself and
its Laplace transform would take the place of micro-canonical and
canonical partition functions. The importance of the corroboration of
this strict analogy is that there is a one to one correspondence between
each specific renewal processes and a thermal system model
with two fields, like a magnet or one-component fluid. The familiar
case of independent identically distributed events is equivalent to the
ideal gas but correlated event processes associate with more interesting
statistical-mechanical systems. Amongst them we chose to look at the
occurrence of phase transitions along time evolution.
\add[r1]{Recently, based on these developments for renewal processes, we constructed 
  a model for cascades of events with application to pandemic spreading under population confinement.}

\subsection{Random walks and fluid correlations}
\label{\myss}

\NewT{
The ingredients for this set of studies originate from two different
sources. One of them the random walk generating functions for the
probabilities of a walker position after a given number of steps. For
symmetric nearest-neighbor step walks on infinite regular lattices
these expressions were known \cite{Montroll2} for several lattice
dimensions and geometries. The other constituent was the
Ornstein-Zernike equation relating pair correlation functions, an
equation that \add[r1]{to be useful} is necessary to \change[r1]{complement}{be complemented} by approximate closure
relations such as the Percus-Yevick approximation \cite{Percus1}. Both
expressions involve convolutions. A general analogy between the two
was developed [\ref{I-i-1}, \ref{I-i-2}] \change[r1]{to}{and was} subsequently
\change[r1]{consider}{used to study} hard-core
lattice gases and the lattice version of the van der Waals model. The
outcome was a description of order-disorder transitions [\ref{I-i-3}], 
decay of fluid correlations [\ref{I-i-4}], an extension to nonequilibrium of the
Ornstein-Zernike relation [\ref{I-i-5}], and a random walk version of phase
coexistence [\ref{I-i-6}]. Independence of events, linearity and lattice
regularity were (and are) the limitations of this approach. There is
no parallel version or development of this unusual analogy between
stochastic processes and equilibrium statistical mechanics, and has
remained basically unnoticed except for sporadic readers in recent
years.
}

\Abs{1}

\Abs{2}

\Abs{3}

\Abs{4}

\Abs{5}

\Abs{6}

\subsection{Random walks and electronic band structures.}
\label{\myss}

\NewT{
As in the previous case two different sources were used to build an
analogy and \add[r1]{derive} from it \remove[r1]{derive} original results. One of them was again the
mentioned random-walk generating function \cite{Percus1}, and the other
the tight-binding Green function equation for electronic band
structure on regular lattices \cite{Mills1}. Again, the two central
expressions for two different physical systems had in common a
convolution, and this was used to make \change[r1]{an}{the} analogy concrete
[\ref{I-ii-1}, \ref{I-ii-2}].
The known trapping procedure to determine first passage times
\cite{Montroll2b} lead to electronic band structures for surfaces and thin
films. To our knowledge there is no parallel version or development of
this analogy. Again, independence of events, linearity and lattice
regularity were (and are) the limitations of this approach.
}

\Abs{8}

\Abs{10}

\subsection{First passage times for multiple walkers.}
\label{\myss}

\NewT{
This development [\ref{I-iii-1}] anteceded similar studies
\cite{Shuler1}. Instead of the common setting of one random walker and one trap, we
considered the successive trapping of an arbitrary number of
independent walkers placed on a lattice each with different given initial
position. The lattice contained also an arbitrary number of given
sites that acted as irreversible traps. The literature at the time
reported modeling the initial events of photosynthesis via trapping of
random walkers. But the sites relevant for the triggering of the
next, chemical, stage required the trapping
of two walkers. Montroll's generating function method \cite{Montroll1} was
generalized to multiple walkers for both basic walks and continuous time walks
[\ref{I-iii-1}]. Independence of walkers, of walker steps, linearity and lattice
regularity were (and are) the limitations of this approach.
}

\Abs{7}

\subsection{Renormalization group and entropy for Weierstrass walks.}
\label{\myss}

\NewT{
Over the years we carried out our statistical-mechanics studies, which
included the consideration of phase transitions, mainly for
equilibrium states, and \change[CV]{this in turn}{mainly} under mean-field
approximations. So, we were disconnected for a long time \remove[r1]{period} from the
mainstream developments in critical phenomena and other topics
involving scale invariance. When we finally considered the
Renormalization Group (RG) method [\ref{I-iv-1}] it was not with any
specific application in mind, but instead we were interested in
exploring its foundations. The RG method appeared to be a procedure
that required devising a specific transformation for each application
\cite{RenormalizationGroup1}. The RG transformation leads to flows away from nontrivial
and towards trivial fixed points. This suggested the monotonic
variation of a function akin to a thermodynamic potential, like
entropy, as the nature of the system changed under the RG
transformation. We looked for an optimization principle as the missing
piece. To this end we selected the Weierstrass random walk \cite{Montroll3}
as a suitable model to investigate this proposition. The scaling
properties of the single step distribution of this walk had been \add[CV]{previously} found to have
similarities with the scaling of thermodynamic properties at
criticality \cite{Montroll3}. We generalized the Weierstrass walk to an
infinite family of related walks with long-tailed single-step
distributions such that the original walk was their RG nontrivial fixed point and
corroborated that the entropy of the step distribution complied with
functional optimization with the fixed points being entropy extrema. Years
later we corroborated this significant feature of the RG method for
network systems
[\ref{III-viii-1}, \ref{III-viii-2}, \ref{III-viii-4}, \ref{III-viii-5}, \ref{III-viii-7}].
To our knowledge there is no similar development of this property.
}

\Abs{109}

\Abs{110}

\Abs{112}

\Abs{115}

\Abs{123}

\subsection{Thermal-system statistical-mechanical analogy for renewal processes.}
\label{\myss}

\NewT{
We considered Montroll's \cite{Montroll2} representation of the generating
function approach for renewal processes and observed that this
formalism is analogous to the statistical-mechanical description of
non-interacting system with two fields, temperature and external
field (or chemical potential) [\ref{I-iv-1}, \ref{I-iv-2}]. The generating function
approach allows for the identification of four different ensembles and
their partition functions (microcanonical, canonical, grand canonical,
and the fourth ensemble that completes the set of conjugate variables
for the Euler relation). Events and time represent degrees of freedom
and energy, respectively. The analogy is straightforwardly extended
to renewal processes with correlated events and, as
expected, the concept of a waiting time (between events) distribution
is no longer useful. But to each two-field thermal system corresponds
a specific renewal event process.  As a mathematical curiosity a
renewal process consisting of \change[r1]{q}{$q$}-independent events was analyzed, where
the normal product of Laplace transformed waiting time distributions
is replaced by the so-called \change[r1]{q}{$q$}-product [\ref{I-iv-1}].
}

\Abs{155}

\Abs{160}

\subsection{Phase transitions along time in correlated renewal processes.}
\label{\myss}

\NewT{
We made specific use of the above-referred renewal process \change[CV]{framework}{approach}
that we \change[CV]{demonstrated}{had shown} to be formally equivalent to the framework formed
by the set of classical ensembles of equilibrium statistical mechanics
[\ref{I-iv-2}]. Most attention has been given to renewal processes of
identical independently distributed (iid) events, but our approach can
equally consider correlated events. Since each specific renewal event
process can be translated into its equivalent (two-field)
statistical-mechanical model system and vice versa, we chose a
well-known solvable statistical-mechanical model exhibiting a
continuous phase transition, the Hamiltonian Mean Field (HMF) model
\cite{Ruffo1, Ruffo2}, and we \change[CV]{use}{made use of} it to describe a phase transition
occurring along time.
Recently, we made use of this exact analogy to develop an original
  stochastic approach for cascading identical events such as contagions 
  in an epidemic or pandemic with application to the COVID 19 contingency. The cascade 
  is made of time shifted renewal event chains such that \remove[CV]{the} its properties are derived 
from those of a renewal process, \change[CV]{that}{and as stated is} in turn \remove[CV]{is} equivalent to a statistical-mechanical thermal 
system. We chose the HMF model and its phase transition to describe the effect of population 
confinement on the spread of contagions.
}

\Abs{155}

\Abs{186}

\newpage

\section{Density Functional Theory}\label{s-DensityFT}  

\textbf{Inhomogeneous states.} Benjamin Widom considered the configurational
(statistical mechanical) effect of the addition of a degree of freedom
in an equilibrium system already composed of many of them \cite{Widom1}. He
found a route, amenable for the study of inhomogeneous systems \cite{Widom2},
to evaluate the equation of state for the chemical potential. We found that
Widom's expression is the Euler-Lagrange equation (the vanishing of
the first variation) associated with the corresponding free energy
density functional. This identification opened many possible studies
of equilibrium states, particularly non-uniform ones, like spatially
ordered states, e.g., solids, phase coexistence, surfaces, etc. The
additional degree of freedom could be a particle, a spin flip,
different types of molecules in a mixture, even a step in a walk, an event
in a renewal process, and so on.

Through this statistical-mechanical route we obtained descriptions for:
liquid to solid transitions in hard core systems, like hard spheres;
interfacial structures for the known (van der Waals) types of binary
mixtures; kinetics of phase change, growth via interface
advance, nucleation and spinodal decomposition. We determined global
phase diagrams for wetting transitions at fluid mixture interfaces,
and developed lattice models for micellar solutions, block copolymers
and microemulsions. We also analyzed the modifications undergone by
surface phase transitions through capillary confinement, including nematics
and quiral molecule mixtures. In relation to wetting transitions we
explored the features of a different type of inhomogeneity, that when
interfaces meet and produces a line of tension. Finally, a particularly
extensive set of studies contain the global phase diagram for the
antiferromagnetic (mean field) spin-1 model, that complements the rich phase
behavior of the (mean field) ferromagnetic case studied previously by Griffiths.

Another formal advance we carried out within density functional theory
was to extend the customary approach to study a planar
interface to a full three-dimensional inhomogeneity. This effort yielded
statistical mechanical expressions for the rigidities of curved
interfaces that are now added to their counterpart for the surface tension,
the so-called Triezenberg-Zwanzig expression derived a couple of
decades earlier. Furthermore, this development made contact with and
gave microscopic support to the phenomenological free energy
expression of Helfrich, a basic starting expression for the study of
microemulsions. Also we extended our approach to the already
mentioned line tension and in general to the stress tensor associated with
deformations descriptive of more general inhomogeneous systems.

This second development led to an additional set of more specific
results, amongst them the suggestion of a new kind of two-dimensional
phase transitions, mediated by the interfacial curvatures of
interfaces. For instance, the transformation of a singly-connected
surface into either multiply-disconnected or multiply-connected
surfaces. The interfacial rigidity appears as a relevant field in a
free energy quantified via the genus of the surface and the
Gauss-Bonnet theorem.  It indicates a possible mechanism for the
inception of micellar solutions or bicontinuous microemulsions from
saturated amphiphile monolayers. Another case arises when considering
the arrest, mediated by the presence of amphiphiles of a phase
separation process, such as nucleation or spinodal decomposition. This
stoppage offers another view of microemulsion formation. Finally, an
extension of the capillary wave model for interfacial curvature fluctuations
followed too.

A third stage was to focus on the basic procedure of the density
functional approach. This consists of two steps, the vanishing of
first variation of the free energy functional and the calculation of the
sign of its second variation fixed by the solution of the 1st condition. The
former locates the stationary solutions while the latter discriminates
between maxima, minima or inflexion-point solutions. As it had been
earlier noticed by Widom \cite{Widom3}, the 1st condition is equivalent to
a classical-mechanical particle problem, while the 2nd condition can
be expressed as a quantum-mechanical particle eigenfunction problem.
In our work these mechanical analogies were used to identify
statistical-mechanical instabilities in simple liquids: the irrelevance
for the position of the Gibbs dividing surface for planar interfaces, the
Raleigh breakup of cylindrical jets, and the Laplace equation at the
onset of spherical droplet nucleation. Other more complex behaviors
were obtained for complex fluids containing amphiphiles.

\change[r1]{Finally}{Additionaly}, we made use of our statistical-mechanical know-how in
relation to the discovery of High $T_c$ superconductivity and the frenzy
of research activity that took place after this. From the body of
accumulated experimental evidence a detailed picture was put together
of a characteristic phase diagram for these materials. A rich assembly
of properties that involves structural phase transitions driven by
small variations of impurities or vacancies, that in turn triggers other phase
changes, from antiferromagnetic to conducting, to superconducting
phases. And to add interest, the mystery of the pairing mechanism. We
contributed with a plausible complete description for the different
phase behaviors and a novel proposition for a pairing mechanism based
on the two-dimensional topological Kosterlitz-Thouless class of phase
transitions \cite{Kosterlitz-Thouless1}.

\add[r1]{Finally, as we describe below in Part IV, recently we have made
  use of density functional theory, in the form of a discrete-time
  Landau equation, to link this subject to low-dimensional nonlinear
  dynamics, their RG fixed-point maps for the routes to chaos, and the
  understanding of the generalized statistical mechanics represented
  by the Tsallis entropy expression.}

\subsection{Density functional theory and Widom's particle insertion method.}
\label{\myss}

\NewT{
In the late Seventies interest appeared in the study of inhomogeneous
equilibrium states \cite{Evans1, RowlinsonWidom1}, such as fluid-wall and
fluid-fluid interfaces, and the phase transition from a fluid to a
solid \cite{Rosenfeld1}. \change[r1]{There}{Two approaches} emerged\remove[CV]{two approaches}, density functional
theory \cite{Evans1, RowlinsonWidom1} and Widom's particle insertion method
\cite{Widom1,Widom2}.
Our early interests focused on exhibiting the basic link
between them. It was shown that Widom's particle insertion formula is
equivalent to the Euler-Lagrange equation, the vanishing of the first
variation of the free energy density [\ref{II-i-1}].
To our knowledge there
is no other development of this equivalency. Widom's formula and \change[r1]{its}{the corresponding}
density functional were then determined for the Kac potential in the
long range, van der Waals limit (a Hamiltonian route for mean field
approximation) and applied to obtain global properties for binary
fluid mixtures [\ref{II-i-2}--\ref{II-i-4}].
}

\Abs{13}

\Abs{12}

\Abs{16}

\Abs{59}

\subsection{Liquid to solid transitions of hard-core model systems.}
\label{\myss}

\NewT{
Also in the late Seventies, in an attempt to `derive' from first
principles the liquid-solid transition in hard-core fluids we looked
first at the simplest possible model, a lattice gas with
nearest-neighbor exclusion, and applied to it a method (our own \remove[r1]{to}
implement\add[r1]{ation of} Widom's formula) that many years later we realized was exact
for a Bethe lattice, that is, the mean field approximation! in a
version suitable for hard core interactions. The next stage was to
solve the problem for an extended hard-core up to an unspecified
number of $n$th-order neighbors. The third stage was to obtain the
continuum space limit by allowing $n$ go to infinity while the lattice
spacing went to zero. The result is a family of Bethe lattice models
that exhibit `liquid to solid' transitions and that are solved
exactly. The main finding for hard spheres is the occurrence of a
bifurcation from the uniform liquid solution at (random) close packing
to a periodic solid-like density. Following the periodic solution from
the bifurcation decreases its density until it can reverse its course
and increase. A density can be determined such that there is a first
order transition before close packing. Thus, the uniform liquid branch
turns metastable at the liquid-solid transition. See Ref. [\ref{II-ii-1}].
 Our derivation of the exact solutions for correlations of
the finite system of hard rods appeared in Ref. [\ref{II-ii-2}]. Years later the
family of Bethe lattice hard-core exclusion models was revisited and
their properties described in further detail [\ref{II-ii-3}]. There
have been many density functional studies of the liquid-solid
transition for hard sphere systems over the years but they do not
encounter or consider the bifurcation at random or uniform close packing \cite{Tarazona1}.
}
  
\Abs{11} 

\Abs{34} 

\Abs{72} 

\subsection{Kinetics of phase change, nucleation, spinodal decomposition.}
\label{\myss}

\NewT{
The Landau\change[r1]{,}{ and the} Cahn-Hilliard \change[r1]{differential}{kinetic} equations \cite{Landau1, Cahn1, Cahn2},
together with the Metiu-Kitahara-Ross extension \cite{Metiu1, Metiu2, Metiu3},
had been previously applied to describe kinetics of phase change in fluid
models under the widely employed square-gradient approximation (the
Landau density functional suitable for slowly varying conditions
near critical points). Our intervention on this topic
involved the use of a more general density functional that we obtained from
our experience with Widom's insertion method. This functional
introduces spatial non-locality and revealed [\ref{II-iii-1}] previously
unknown features, such as: i) Cooperation of non-locally correlated
fluctuations that provide the nucleation process with an additional
term absent in the classical and gradient theories. ii) And an
infinite family of periodic stationary states within the spinodal
region that serve as a ladder to reach stability in the approach to equilibrium.
Our advance opened the way for: the determination of similar
features in segregating and ordering alloys [\ref{II-iii-2}]; an extension
of irreversible thermodynamics [\ref{II-iii-3}]; the observation of the effect of
confinement on spinodal decomposition [\ref{II-iii-4}]; and the influence of
wetting on kinetics of phase change in fluid mixtures [\ref{II-iii-5}].
}

\Abs{14} 

\Abs{15} 

\Abs{37} 

\Abs{99} 

\Abs{106} 

\subsection{Global wetting phase diagram for fluid interfaces.}
\label{\myss}

\NewT{
Our contributions were not the earliest wetting transition
developments, mostly done for planar solid to fluid interfaces
\cite{Wetting1, Wetting2, Wetting3, Cahn3, Ebner1},
but the first dealing with fluid-fluid interfaces. We
constructed a global phase diagram for the entire set of interfaces of the van der Waals
binary fluid mixture types [\ref{II-iv-1}] that is the counterpart for the
global phase diagram for planar solid to fluid interfaces
\cite{NakanishiFisher1}. The methodology was mostly based on a non-local
density functional that generalized the common slowly varying square
gradient approximation. We addressed some unusual systems that did not
receive attention in other studies of wetting, such as the wetting
properties of model semipermeable membranes [\ref{II-iv-3},\ref{II-iv-4}], pinning of alloy
anti-phase boundaries [\ref{II-iv-5}], and the effect of wetting on the kinetics
of phase change [\ref{II-iv-8}]. We also studied experimentally some related
phenomena, such as the possible presence of prewetting in transient
foams [\ref{II-iv-6}], and of a continuous wetting transition in a specific fluid
mixture [\ref{II-iv-7}].
}

\Abs{18} 

\Abs{19}

\Abs{26} 

\Abs{27} 

\Abs{28}

\Abs{32} 

\Abs{48} 

\Abs{106} 

\subsection{Lattice models for micellar solutions, microemulsions, magnetic alloys, etc.}
\label{\myss}

\NewT{
We made an attempt to contend with the teams seeking to develop a
lattice model to describe the interesting microemulsion phase, and the
transitions it undergoes. Generalizations of the original
Widom-Wheeler model \cite{WidomWheeler1, WidomWheeler2} were developed to address
different types of systems. The first of them was designed to be
equivalent to the solid-on-solid Ising model that was found to display
the elusive roughening transition \cite{Weeks1}. Through an exact analogy
a representation of the disordering of a lamellar diblock polymer
microemulsion was obtained [\ref{II-v-1}]. A second generalization [\ref{II-v-2}] to an
additional type of molecular end (of the two original types) that form
the bifunctional molecules, the Widom-Wheeler mixture model expanded in such a way as
to be exactly analogous to the famous Griffiths' three-component model
(the mean-field spin-1 model) \cite{Griffiths1}. A third generalization
[\ref{II-v-3}--\ref{II-v-6}] to finite interactions of the original Widom-Wheeler model led
not only to a rich global micellar solution model phase behavior, but
also (again but in a different way) to an analog of the
symmetric portion of the Griffiths three-component model with
staggered field [\ref{II-v-5}]. Finally, we extended Griffiths original work, a
vast global phase diagram for uniform, or ferromagnetic, phases to
sublattice ordered, or antiferromagnetic, phases [\ref{II-v-7}]. New applications
appeared for ordering magnetic alloys [\ref{II-v-8}] and to micellar solution
models [\ref{II-v-9}]. The most important contribution [\ref{II-v-7}] was revealing a rich
phase behavior where ordered phases are present that had remained
hidden under the known global phase diagram for uniform phases made by
Griffiths and coworkers. The global phase diagrams in
Refs. \cite{Griffiths1} and [\ref{II-v-7}] are like the two different faces of the
moon.
}

\Abs{29} 

\Abs{31} 

\Abs{33} 

\Abs{38} 

\Abs{39}

\Abs{50} 

\Abs{53} 

\Abs{54} 

\Abs{55} 

\subsection{Anomalous micellar solubility loops.}
\label{\myss}

\NewT{
Nonionic amphiphile aqueous solutions exhibit asymmetric solubility
loops and experimental determination of critical exponents at their
lower critical solubility point revealed anomalous behavior suggesting
nonuniversality of the critical indexes \cite{Anomalous1, Anomalous2}. The extension of
the Widom-Wheeler model we had made to finite interactions and its
analogue as a micellar solution model became a suitable platform to
obtain, first of all, solubility loops [\ref{II-vi-1},\ref{II-vi-2}], and then, the more
interesting and experimentally available, asymmetric solubility loops
[\ref{II-vi-3},\ref{II-vi-4}]. We therefore had the opportunity to investigate theoretically
the experimentally detected critical behavior and address the polemical
observations. The model is equivalent to a temperature and external
field dependent spin-1/2 Ising magnet, and this feature provides a
rationalization that explains the controversial experimental
observations, that in turn led to explicit approbatory reference
\cite{Goldstein1}. Our model was suitable for further studies such as
micellar interfacial and capillary properties [\ref{II-vi-5}] that resulted in the
characterization of enhanced amphiphile adsorption at their
liquid-liquid interfaces [\ref{II-vi-6}].
}

\Abs{57} 

\Abs{63} 

\Abs{67} 

\Abs{71} 

\Abs{97} 

\Abs{104} 

\subsection{Complex fluids under confinement.}
\label{\myss}

\NewT{
Density functional theory proved to be a suitable method for the study
of the finite size effects on the properties of inhomogeneous
systems. Phase transitions are modified when a three-dimensional
system is confined in size along only one spatial dimension
\cite{Confinement1}. Some years after the topic received the first wave of
attention by the statistical physics community we
analyzed a few new situations. Some of them correspond to the same
model studies that ante-ceded us but aimed at gaining new knowledge
from them. We studied the modification undergone by spinodal
decomposition under confinement and observed the development of
lamellar or strip patterns [\ref{II-vii-1}]. Other works aimed at the description
of phase transitions induced or suppressed by confinement [\ref{II-vii-6}] that
become relevant when the models represent nematic liquids [\ref{II-vii-2}] or
enantiomeric mixtures [\ref{II-vii-5}].  \change[r1]{We also}{Additionaly, we} put together a global description
for the effects of confinement through the family of surface phase transitions
exhibited by a semi-infinite model magnet or simple fluid [\ref{II-vii-3},\ref{II-vii-7}].
We also determined density fluctuations and correlations when a model system
becomes finite \change[CV]{in}{along} one dimension [\ref{II-vii-4}].
}

\Abs{99} 

\Abs{101} 

\Abs{105} 

\Abs{107} 

\Abs{113} 

\Abs{116} 

\Abs{118} 

\subsection{Curved interfaces, bending rigidities, line tension, stress tensor and capillary waves.}
\label{\myss}

\NewT{
We considered density functional theory for inhomogeneous systems for
which surface tension is not the only important or the main free
energy contribution, and determined bending constants, interfacial
width and line tension [\ref{II-viii-1}].  In relation to the microemulsion problem
we went smoothly from lattice models, and studied the properties of
curved interfaces in continuum space, of particular importance because
of the vanishing of the surface tension.
\change[CV]{We}{From a `microscopic' starting  point we} derived 
expressions for the bending terms that are `mesoscopic' quantities,
going one step beyond the usual surface tension description. Indeed,
by considering cylindrical, spherical, and then general surface shape
fluctuations it was possible to derive rigorous expressions for the
bending constants in terms of pair correlations [\ref{II-viii-2},\ref{II-viii-3}], similar to the
expression of Triezenberg and Zwanzig for the surface tension obtained
many years before. Amongst other results we, and collaborators, were
able to generalize the capillary-wave model of an interface and
explain its properties [\ref{II-viii-4},\ref{II-viii-7},\ref{II-viii-11},\ref{II-viii-12}].
We also derived explicit
expressions for the stress tensor for general inhomogeneities in a
one-component simple fluid in terms of density gradients and moments
of the direct correlation function [\ref{II-viii-5},\ref{II-viii-9}]. We were the first with these
closed-form expressions, and with the conceptual clarification of the
phenomenological Helfrich free energy that governs the physics of
curved interfaces [\ref{II-viii-6},\ref{II-viii-8},\ref{II-viii-10},\ref{II-viii-13}].
Our results contributed to provide the
statistical-mechanical basis for the free energy terms of curved
interfaces originally derived from phenomenological elasticity
arguments.
}

\Abs{73} 

\Abs{74} 

\Abs{75} 

\Abs{80} 

\Abs{86} 

\Abs{91} 

\Abs{92}

\Abs{95} 

\Abs{96} 

\Abs{98} 

\Abs{100} 


\Abs{111} 

\Abs{114} 

\subsection{Curvature interfacial transitions.}
\label{\myss}

\NewT{
As a consequence of our understanding of the Helfrich free energy of
curved interfaces we were in a position to suggest the possibility of
novel types of surface phase transitions that could take place in
thin interfaces like \change[CV]{are}{such as} amphiphile interfaces.
The first \add[CV]{new} kind of
transition corresponds to multi-patch buckling of the surface and its
associated order parameter is its mean curvature [\ref{II-ix-1}--\ref{II-ix-3}].
The second, topologically-driven, \add[CV]{novel} transition transforms a simply-connected
surface state into a multiply disconnected (or multiply connected)
volume-filling surface state. Its order parameter is given by the
surface genus (or number of micelles). These volume-filling surface
states arise when the free energy cost for the creation of a droplet
or a `handle' out of the original monolayer surface vanishes. Our
results follow from the Gauss-Bonnet theorem that links local
curvature properties with the surface global topological invariants
[\ref{II-ix-1}--\ref{II-ix-3}]. Finally, we made use of the results from the
Helfrich free energy studies to obtain an interpretation of the phase
properties of microemulsions as a detained\add[r1]{, or arrested,} ordinary phase separation
due to the action of amphiphiles [\ref{II-ix-4}]. That is, we put forward
an interpretation of the characteristic phase properties of microemulsions
based on a distinct process of phase separation that has come to a
stand-still due to the action of the amphiphiles on the driving force
of the process. In our kinetic equation this force relates to the generalized
Laplace equation that contains spontaneous curvature and bending
rigidity terms, and, according to it, the final stationary states attained
are phases structured into water-rich and oil-rich domains [\ref{II-ix-4}].
}

\Abs{66} 

\Abs{70} 

\Abs{81}

\Abs{103}    

\subsection{Line tension \& wetting.}
\label{\myss}

\NewT{
In parallel with these developments we looked at a different type of
inhomogeneity, the macroscopic line that is formed by the intersection
of three interfaces and that carries a free energy cost, the line of
tension. First: we considered the change in grand potential due to a
line inhomogeneity where two or more interfaces meet, and derive an
exact expression for its tension in terms of the direct correlation
function and the gradients of the densities. We proposed a model where
a line inhomogeneity is formed at the intersection of the free surface
and a domain boundary of an Ising magnet [\ref{II-x-1},\ref{II-x-4}]. Second: we calculated,
for a spin-1 Ising model within the mean-field approximation, the line
tension along partial-wetting surface states up to a first-order
wetting transition where the line disappears. We likewise calculated
the line tension of the boundary between the two coexisting surface
states at the prewetting transition and follow its behavior into the
neighborhood of bulk wetting. In both cases we find evidence for the
divergence of the line tension [\ref{II-x-2}]. Third: we extended the interfacial
wetting transition to a similar phenomenon for the contact line
together with an Antonov rule for the line tension [\ref{II-x-3}]. The standard
wetting transition consists of the transformation of a microscopically
thin two-dimensional interface into a macroscopically thick structure
composed of two interfaces separated by a bulk phase. We considered
the one-dimensional analog of this phenomenon, when a contact line
among three or more phases decomposes into two contact lines separated
by an interface. Our findings help both in settling the discussion on
the limiting value of the line tension and in understanding the origin
of its singular behavior. Fourth: we studied the effect of
fluctuations on the line tension at first-order and continuous wetting
transitions. We considered thermal wandering, strong fluctuations as
in random media, and weak fluctuations as in quasiperiodic
systems. These properties establish the occurrence of hyperscaling and
nonclassical exponents for the line tension at wetting. [\ref{II-x-6},\ref{II-x-8}]. Fifth:
we studied the three-phase contact line and its tension near the
interfacial phase transition from partial to complete wetting. We
analyzed the singularity of the line tension at first-order wetting
transitions and showed that it displays the universal features of a
critical endpoint [\ref{II-x-7}].
}

\Abs{76} 

\Abs{77} 

\Abs{78} 

\Abs{79} 

\Abs{83} 

\Abs{84} 

\Abs{87} 

\Abs{89} 

\subsection{Analogy of density functional 1st \& 2nd variations with classical and quantum mechanics.}
\label{\myss}

\NewT{
We studied first the stability of cylindrical and spherical interfaces
with respect to density fluctuations within the square-gradient
approximation [\ref{II-xi-1}]. That is, we determined the stability matrix (of the
second derivatives of the free energy functional with respect to the
density) when the stationary state is a cylindrical or spherical
droplet of a stable phase embedded in the metastable phase. For these
geometries the stationary states are unstable, some of the eigenvalues
are negative and their eigenfunctions represent those fluctuations
that are amplified in a process where the equilibrium state is
reached. At early times, in a simple Landau kinetics model, the
eigenfunctions represent the fluctuations that grow or decay with a
simple exponential law and with a characteristic time that is
proportional to the inverse of the eigenvalues. Our results agree with
the stability criteria obtained from the Laplace equation, that is,
the nucleation of critical droplets and in the case of cylinders also
the Rayleigh instability. In the limit of infinite radius we recover
the known results for the planar interface between two stable
phases. The modes with lowest energy correspond to the customary
capillary waves, while other modes with higher energy associated to
changes in the interfacial width are shown to be related to a novel
interfacial coefficient. \change[CV]{Then}{Subsequently,} we carried out a comprehensive
description of interfaces containing amphiphiles developed through the
use of a free energy density functional with squared-gradient and
squared-Laplacian terms [\ref{II-xi-2},\ref{II-xi-3}]. This elemental model functional
contains the basic ingredients to examine interfacial stability, it is
technically tractable and a range of results for curved interfaces,
many in analytical form, have been obtained from it. These are: (i)
average equilibrium properties, such as pressure tensor, interfacial
tension and position of the Gibbs dividing surface, (ii) order
parameter fluctuation modes and stability matrix, and, (iii) an
effective interfacial potential that in the small curvature limit
corresponds to the Helfrich free energy. Our survey was meant to fill
an existing gap, mostly conceptual, left by previous work.
}

\Abs{102} 

\Abs{111} 

\Abs{117} 

\subsection{Phase behavior and pairing mechanism for two-dimensional superconductors.}
\label{\myss}

\NewT{
We studied the properties of the copper oxide High-$T_c$ superconductors
from a statistical-mechanical perspective. First, we developed a model
for twin boundaries, like those observed in YBCO ceramic
superconductors. The model, based on an inhomogeneous Landau-Ginsburg
free energy, was used to show that the twin boundaries induce a purely
interfacial superconducting transition at a temperature above the bulk
transition temperature [\ref{II-xii-1},\ref{II-xii-3}]. Second, we developed a model for the
absorption of oxygen in the copper-oxide basal planes of YBCO ceramic
superconductors based on a layered oxygen lattice gas model in
equilibrium with an external oxygen source. We considered, in addition
to Cu-mediated oxygen-oxygen interactions, the elastic energy of the
crystal. The model exhibits coupled oxygen ordering and tetragonal to
orthorhombic structural transitions. We obtained quantitative
agreement with experimental data [\ref{II-xii-2},\ref{II-xii-4},\ref{II-xii-5}]. Third, we developed a
spin-hole model for the magnetic phase behavior of weakly coupled
copper-oxygen layers. For the hole-free system there are
antiferromagnetic couplings between Cu magnetic moments and weak Ising
anisotropy with canting. Addition of holes, localized on the O sites,
induces a transformation of anisotropy in the spin couplings, from
Ising to XY, and of its sign, from antiferromagnetic to ferromagnetic
[\ref{II-xii-6},\ref{II-xii-7}]. Fourth, the spin-hole model was developed further via the use
of the Moriya-Dzyaloshinsky Hamiltonian for anisotropic Heisenberg
spin couplings. The global phase diagram for the ceramic
superconductors became better rationalized. Addition of holes (via
doping) smooths corrugation of the copper-oxygen planes, that in turn
drives the structural phase transition from tetragonal to orthogonal, and
changes the magnetic behavior from Ising-like antiferromagnetic
with a sharp Neel temperature to XY-like with weak ferromagnetic
couplings [\ref{II-xii-8},\ref{II-xii-9}]. Fifth, the spin-hole model implies a novel kind of
charge pairing mechanism for the copper oxide High Tc
superconductors. The vortex magnetic excitations occurring in the
copper-oxide planes capture mobile charges or holes, and the pairing
of these occurs via the Kosterlitz-Thouless (KT) vortex-antivortex
pairing mechanism. [\ref{II-xii-7}--\ref{II-xii-9}]. Sixth, the phase properties of the
spin-hole model were found to be consistent with an extension of the
Hubbard Hamiltonian model to competing positive-and negative-$U$
interactions on a 2D lattice. The positive-$U$ Hubbard model exhibits
Ising antiferromagnetic phase behavior, while the negative-$U$ Hubbard
model displays XY superconductivity. The phase progression observed in
High-$T_c$ superconductors, and in the spin-hole model, can be obtained
via gradual change in the concentration of initially purely $U>0$
interactions via addition of $U<0$ interactions, like in a binary alloy
[\ref{II-xii-10}]. It is only recently that the Moriya-Dzyaloshinsky interaction
and the KT transition has been associated with the properties of
copper oxide High-$T_c$ superconductors \cite{Luca1,Konig1}.
}

\Abs{40} 

\Abs{41} 

\Abs{43} 

\Abs{44} 

\Abs{46} 

\Abs{49} 

\Abs{56} 

\Abs{58} 

\Abs{64} 

\Abs{88} 

\newpage

\section{Nonlinear Dynamics}\label{s-NonlinDyn}  

\textbf{Onset of chaos.} A chance communication between Robert May
and James Yorke in 1973 \cite{SanJuan1} led to the comprehension of the
bifurcation diagram of the quadratic map, perhaps nowadays the better-known
picture symbolizing nonlinear dynamics. This diagram illustrates the
families of attractors or stable solutions for this difference
equation and was first put together by linking the works of May \cite{May1}
and Yorke and Tien-Yien-Li \cite{Yorke1}. Not long after this, Feigengaum
contributed \cite{Feigenbaum1,Feigenbaum2} to the full understanding of the universal
scaling properties at the period-doubling accumulation point, the
borderline between regular and irregular behaviors, the transition in
and out of chaos.

Twenty\add[r1]{-odd} years ago we began, then increased and more recently
consolidated, a linked set of research lines dedicated to the
investigation of complex systems. This set rests on independent,
original and rigorous developments on the transitions to chaos in
dissipative nonlinear systems of low dimensionality, among which the
quadratic map has been the most frequently used. Our investigations
are based on two groups of properties we obtained and developed
into effective research tools: the dynamics inside and the dynamics
towards the attractors that represent the transitions to chaos. Based
on this knowledge, studies on core problems of complex systems
have been carried out. In the physics of condensed matter we have
obtained results in problems of difficult treatment: glassy dynamics,
localization and critical fluctuations. In complex systems we have
achieved central results for evolutionary dynamics, ranked data
distributions (Zipf and Benford laws), and\remove[CV]{, recently,} on the
rationalization and (needed) statistical-mechanical justification of
the phenomenon of self-organization\change[CV]{, and recently on}{. Recently we demonstrated}
the equivalence
of paradigms in this field, edge of chaos and criticality.}

Our motivation to examine the transitions to chaos displayed by
  prototypical nonlinear iterated maps, such as the logistic and
  circle maps, originated from the goal \change[CV]{of}{to} understand\remove[CV]{ing} the possible
  existence of a limit of validity of the ordinary Boltzmann-Gibbs
  statistical mechanics and the conceivable appropriateness of the
  alternative entropy expression proposed by Tsallis \remove[CV]{[Ref]} [\ref{IV-ii-1}].
  The
  failure of both ergodic and mixing properties at these transitions,
  valid for their neighboring chaotic attractors pointed at a feasible
  ‘numerical laboratory’ in which to explore these issues. As a
  consequence of this all the studies described in this Part III
  contain results related to our quest of comprehending the so-called
  $q$-statistics.
  
With regards to complex
condensed matter physics problems we mention the following: The joint
use of density functional theory (for inhomogeneous systems) and the
renormalization group fixed-point map (for the tangent bifurcation)
to deliver a space-time description for the dominant fluctuation at a
critical point. The bifurcation gap induced by addition of external
noise, that removes the period-doubling and chaotic-band-splitting
accumulation points, was shown to appear as a crossover phenomenon
along time evolution at the onset of chaos. We found that before the
crossover the dynamics is analogous to glassy dynamics in molecular
systems. In addition a vital recursion relation for size growth of a basic wave
scattering model was recognized as a nonlinear iteration map with a
bifurcation diagram where tangent bifurcations separate periodic
(insulating) and chaotic (conducting) attractors.

Amongst specific
advances in nonlinear dynamics we make reference to: A central
quantity in nonlinear dynamics, the sensitivity to initial conditions,
was determined explicitly for the pitchfork and tangent bifurcations
as well as for the period doubling and the golden ratio quasi-periodic
transitions to chaos, all of which display anomalous properties while
their Lyapunov exponent vanishes. The distributions for sums of
successive positions of trajectories, as in random walks, for families
of chaotic attractors of the quadratic map were found to conform to a
renormalization group scheme such that its trivial fixed point matches
the central limit theorem. From the start, the property that
summarized our calculations for the dynamics towards the periodic
attractors of the quadratic map resembled \change{in form}{formally} that of a partition
function. Continued work supported this interpretation and finally a
model for self-organization materialized.

With reference to
contributions to the general understanding of complex systems we
annotate: The initial consideration of the discrete-time version of
the replicator equation of established models of evolutionary game
theory lead straightway to novel bifurcation diagrams and valuable
coupled-map models. The transformation of iterated map trajectories
into (Horizontal Visibility) networks opened the possibility of
reevaluating the link of the renormalization group technique with
entropy optimization, and likewise the generalization of the Pesin
theorem at the transitions to chaos. The size-rank distribution of all
kinds of numerical data, including Zipf's law, were seen to be
obtainable from nonlinear maps close to a tangent bifurcation, such
that data samples are reproduced by their trajectories.

For reviews that contain descriptions of these developments
see \cite{Review1, Review2, Review3}.

\subsection{Critical fluctuations and the intermittency route out of chaos.}
\label{\myss}

\NewT{
We followed up studies \cite{Antoniou1, Contoyiannis1} for the
spatial structure and temporal evolution of fluctuations at an
ordinary continuous phase transition. These studies employed the
Landau-Ginzburg-Wilson (LGW) effective free energy together with a
deduced nonlinear iterated map near tangency. Detailed results were
obtained for the dominant fluctuation, a large \add[r1]{and relatively} long-lived object
obtained via the saddle-point approximation. More over, the time
evolution of the dominant fluctuation was found to be of the
intermittent type. The setting was ready for a possible
statistical-mechanical application of the renormalization group
fixed-point map at the tangent bifurcation, which would carry
implications on its generalization known as nonextensive statistical
mechanics. The employment of both the density functional theory (for
inhomogeneous systems) and the renormalization group fixed-point map
(for the tangent bifurcation) delivered a space-time description for
the dominant fluctuation at a critical point [\ref{III-i-2}--\ref{III-i-4}].
Thus, we
established an unforeseen connection between the area of critical
phenomena in statistical mechanics and anomalous nonlinear dynamics at
the transitions to chaos, a setting where generalized $q$-deformed
entropy expressions appear naturally.
}

\Abs{130} 

\Abs{134} 

\Abs{145} 

\Abs{179} 

\subsection{Glassy dynamics at the noise-perturbed onset of chaos.} 
\label{\myss}

\NewT{
The bifurcation gap induced by the addition of external noise
\cite{Schuster1}, that removes the finer features of the         
period-doubling and chaotic-band-splitting cascades together with
their accumulation points, was shown to appear as a crossover
phenomenon along time evolution at the onset of chaos [\ref{III-ii-1}--\ref{III-ii-5}]. It has
been some time since we pointed out an unusual connection between two
seemingly different situations [\ref{III-ii-1},\ref{III-ii-2}], one of them the bifurcation
gap and the other glass formation. The first belongs to nonlinear
dynamics and the second to condensed matter physics. It was shown that
they share main defining properties: the gradual disappearance of
diffusion, the scaling law known as aging, anomalous two-step
relaxation, etc. [\ref{III-ii-1}--\ref{III-ii-5}]. Noise amplitude in one setting represents
temperature distance to a vitreous state in the other. The two
problems indicate loss of ergodicity as noise amplitude or temperature
distance vanishes. The quadratic map with additive noise can be seen
to be a kind of discrete-time nonlinear Langevin
equation. Nonetheless, in one case there is only functional
composition while in the other there are molecular collisions. We have
shown that the ideal glass concept exists and can be precisely
represented by the attractor at the onset of chaos. This requires the
knowledge that the noise-induced bifurcation gap is recapitulated at
such, perturbed, onset of chaos. \add[r1]{Before the crossover glassy dynamics
  is governed by $q$-statistics while after chaotic dynamics is reestablished
  ordinary statistics is recovered.}
}

\Abs{128} 

\Abs{129} 

\Abs{131} 

\Abs{132} 

\Abs{136} 

\subsection{Localization transition as a tangent bifurcation.}
\label{\myss}

\NewT{
A deep running analogy was uncovered between two apparently different
physical problems, which allowed for the determination of elusive
quantities and understanding of difficult issues. Intermittency and
electronic transport were found to share common features and bridged
the fields of research in nonlinear dynamics and condensed-matter
physics. Specifically, the dynamics at the onset of chaos \add[CV]{via intermittency} appears
associated with the critical conductance at the mobility edge of
regular self-similar scatterer networks [\ref{III-iii-1}]. More specifically, wave              
propagation through scattering media described by means of a
double Cayley tree permits full solutions as the Bethe lattice, a form
of mean-field, simplifies sufficiently the \add[r1]{calculation of the} scattering matrix\remove[r1]{ calculation}.
Other physical situations where the localization
phenomenon occurs, light, sound or elastic media wave scattering, can
be likewise modeled and described by nonlinear dynamics of low
dimensionality, with the underlying implication of a drastic reduction
of degrees of freedom [\ref{III-iii-1}--\ref{III-iii-4}]. This dynamics is represented by the 
M\"obius transformation, the fixed points of which correspond to the
localized states, and its ever-changing positions or phases to the
extended states that display coherence due to vanishing Lyapunov
exponent [\ref{III-iii-2}]. Similar reductions of degrees of freedom leading to         
M\"obius transformations have been observed in the synchronization of
arrays of oscillators \cite{Marvel1}.
\add[r1]{The conducting, coherent, weakly-chaotic regime exhibits $q$-statistics.}
}

\Abs{151} 

\Abs{162} 

\Abs{178} 

\Abs{183} 

\subsection{Generalization of the Pesin identity at the Feigenbaum attractor.}
\label{\myss}

\NewT{
A central quantity in nonlinear dynamics, the sensitivity to initial
conditions, was determined explicitly for the first time for the pitchfork and tangent
bifurcations as well as for the period doubling and the golden ratio
quasi-periodic transitions to chaos                                    
[\ref{III-iv-1},\ref{III-iv-2},\ref{III-iv-5}--\ref{III-iv-7},\ref{III-iv-9},\ref{III-iv-10}],
all of which display
anomalous properties while their Lyapunov exponent vanishes. At the
transitions to chaos in one-dimensional nonlinear maps the (only)
Lyapunov exponent vanishes. A classical example is the period-doubling
accumulation point attractor (the Feigenbaum attractor) that
appears an infinite number of times in the bifurcation diagram of
quadratic maps. There, the sensitivity to initial conditions behaves
anomalously; it fluctuates endlessly in iteration time with increasing
amplitude that grows sub exponentially. There is a remarkable property
in the dynamics at the Feigenbaum attractor that leads to an identity
between the $q$-generalized Lyapunov exponent and the rate of growth of
the $q$-generalized entropy. This is the counterpart of the Pesin
identity that states the equality of the (positive) ordinary Lyapunov
exponent with the Sinai-Kolmogorov entropy for chaotic attractors.
 This property is that a distribution, say uniform, of initial 
conditions within a small interval, remains invariant for selected
iteration times. The resulting $q$-generalized Pesin identity [\ref{III-iv-6}] can be
arguably considered to be the most important solid evidence regarding
the so-called $q$-statistics \cite{Tsallis1}.
}

\Abs{119} 

\Abs{120} 

\Abs{121} 

\Abs{123} 

\Abs{124} 

\Abs{125} 

\Abs{135} 

\Abs{139} 

\Abs{143} 

\Abs{144} 

\subsection{Renormalization group and central limit theorem for chaotic attractors.}
\label{\myss}

\NewT{
A claim \cite{Tirnakli1,Tirnakli2,Tirnakli3,Tirnakli4} about a novel kind of central limit stationary
distribution for correlated variables to be displayed at the
period-doubling onset of chaos attracted our attention and these led
us to examine sums of positions of trajectories. This effort led us to
clarify the issue by uncovering a remarkable renormalization group
picture. The distributions for sums of successive positions of
trajectories, as in random walks, for families of chaotic attractors
of the quadratic map were found to conform to a renormalization group
scheme such that its trivial fixed-point matches the central limit
theorem [\ref{III-v-1}--\ref{III-v-5}].
In our work [\ref{III-v-1}--\ref{III-v-5}]                        
we considered sums of positions
from a single trajectory and also from an ensemble of them, the single
trajectory at the transition to chaos leads to a multifractal-valued
sum, that once re-scaled is similar to the trajectory itself. For
ensembles of trajectories the structure of the fractal function, which
is this sum at the onset of chaos as a function of the initial
condition, is built in stages that recapitulate the additional
increasingly finer features added along the period-doubling
cascade. For chaotic band attractors the evolution of the distribution
of sums of positions, of a single trajectory, or of an ensemble of
them, inevitably takes the stationary Gaussian form as the number of
terms diverges. \add[r1]{At the crossover from multifractal to gaussian
behavior the distribution takes the form of a $q$-gaussian.}
}

\Abs{152} 

\Abs{154} 

\Abs{169} 

\Abs{171} 

\Abs{176} 

\subsection{Self-organization along the period-doubling cascade.}
\label{\myss}

\NewT{
The property that summarized our calculations for the dynamics towards
the periodic attractors of the quadratic map resembled in form that of
a partition function. Subsequent work gave support to this
interpretation and finally a model for self-organization materialized
from it [\ref{III-vi-1}--\ref{III-vi-8}].                               
We computed properties of ensembles of trajectories
evolving towards (first periodic and then chaotic) attractors of the
quadratic map. The expression descriptive of the space populated by
trajectory positions at any given iteration time under the action of
the attractor was identified as a partition function, equivalent to
that of the construction by stages of a multifractal set. Ultimately,
this statistical-mechanical likeness became a model for
self-organization. The time evolution of the fraction of occupied
space in logarithmic scales exhibits the telltale occurrence of
discrete scale invariance, power-law decay dressed by logarithmic
oscillations, and displays a `recapitulation' property, i.e. evolution
towards periodic or chaotic-band attractors repeats successively the
evolution towards those attractors with smaller periods or number of
bands. Only recently [\ref{III-vi-7},\ref{III-vi-8}],                
the statistical-mechanical justification
of the fraction of occupied space as a bona fide partition function
was put together. The balance between numbers of configurations and
Boltzmann-Gibbs statistical weights of the initial thermal system is
strongly altered and ultimately eliminated by the sequential
subdivision procedure that mirrors the actions of the
attractor. However, the emerging set of subsystem configurations
implies a different and novel entropy growth process that eventually
upsets the original loss and has the capability of marginally [\ref{III-vi-8}]  
locking the system into a self-organized state with characteristics of
criticality, as in the so called self-organized criticality \cite{Jensen1}.
At the transition to chaos self-organization displays full
scale-invariant properties similar to space and time scale invariance
of critical states. There, the number of subsystem configurations and
their generalized                                            
entropy, a $q$-entropy, is maximal [\ref{III-vi-8}]. Attaining this state
provides an explanation, within perhaps the simplest model system, for
the hypothesis of self-organized criticality \cite{Jensen1}.
}

\Abs{146} 

\Abs{147} 

\Abs{148} 

\Abs{149} 

\Abs{153} 

\Abs{164} 

\Abs{170} 

\Abs{185} 

\subsection{Chaos in discrete-time game theory.}
\label{\myss}

\NewT{
Another case study \add[r1]{we} developed consisted of the inspection of the
consequences of introducing discrete time to the replicator equation
for a collection of well-known (social) games. We were headed
straightforwardly into a nonlinear-dynamical extension of evolutionary
game theory [\ref{III-vii-1}]. We focused attention on the simplest option, the
well-known social-dilemma (two-strategy, cooperation or defection)
games represented by symmetric two-by-two payoff matrices. The
appearance of chaotic dynamics in these games is ruled out by the
Poincare-Bendixon theorem that establishes the requirement of at least
three dimensions for the occurrence of chaos in a continuous time
dynamical system. Therefore we chose to introduce discrete time into
the replicator equation and convert it into a nonlinear iterated map
with two control parameters [\ref{III-vii-1}]. The results were immediate, the
landscape of the three-dimensional bifurcation diagram uncovers a rich
arrangement of periodic and chaotic attractors connected by
recognizable but somewhat distorted period-doubling and chaotic band
splitting cascades, windows of periodicity, etc. [\ref{III-vii-1}]. Our nonlinear
iterated map can be obtained after introduction of approximations to
high-dimensional coupled-map models representative of evolutionary
dynamics of eco\remove[CV]{logical }systems, like the Tangled Nature model \cite{Jensen2}.  
In doing so we have obtained [\ref{III-vii-2}, \ref{III-vii-3}] with this
approximation the possible connection between the
macroscopic intermittent behaviors of the above-mentioned
high-dimensional models with the known low-dimensional sources of
intermittency, such as the tangent bifurcation. \add[r1]{As a result
  $q$-properties at pitchfork and tangent bifurcations and at transitions
  to chaos have analogues in social games.}
}

\Abs{159} 

\Abs{177} 

\Abs{180} 

\subsection{Complex network view of the routes to chaos.}
\label{\myss}

\NewT{
The transformation of iterated map trajectories into (Horizontal
Visibility, HV) graphs opened the possibility of reevaluating the link
of the renormalization group technique with entropy optimization, and
likewise the generalization of the Pesin theorem at the transitions to
chaos [\ref{III-viii-1}--\ref{III-viii-7}].
The examination of their \add[CV]{resultant} network structure\add[CV]{s}, their
degree distribution and their entropy expressions produced another
significant result: the \change[CV]{occurrence}{exposure} of an HV network version of the
$q$-generalized Pesin identity at the period-doubling [\ref{III-viii-3}] and at the
quasi-periodic [\ref{III-viii-6}] transitions to chaos. In both cases, the
`refinement' quality of the HV algorithm, many time series into one
graph, simplifies the multifractal set at the transition to chaos
attractor into a fractal set. This advance required the network
definitions for sensitivity to initial conditions and for Lyapunov
exponent. The structure of the HV networks obtained from attractor
trajectories lends itself in all cases to the consideration of a
simple (contiguous-node-merging) renormalization-group (RG)
transformation [\ref{III-viii-1}, \ref{III-viii-2}, \ref{III-viii-4}, \ref{III-viii-5}].
For the logistic map there are two
trivial fixed-point HV graphs, those from period one (single chain)
and from a single chaotic band (single chain dressed with random
links), and one nontrivial fixed-point (scale-invariant) graph, that
for the transition to chaos [\ref{III-viii-1},\ref{III-viii-2}].
We have pointed out [\ref{I-iv-1}]                                        
that there is a hidden entropy optimization procedure underlying the
renormalization group technique\add[CV]{, useful} for extracting properties of systems
with scale-invariant properties. Specifically, that the all-important
trivial and nontrivial fixed points are extrema of a suitably defined
entropy. The access to entropy is provided via the network degree
distributions through the Shannon expression. \add[r1]{The HV networks
  for period-doubling, quasi-periodicity and intermittency routes to
  chaos were determined.}
}

\Abs{158} 

\Abs{161} 

\Abs{163} 

\Abs{165} 

\Abs{166} 

\Abs{167} 

\Abs{175} 

\subsection{Universality classes of rank distributions revealed by nonlinear maps near tangency.}
\label{\myss}

\NewT{
The consideration of an existing stochastic approach for the
reproduction of ranked data \cite{Pietronero1} led to a formal equivalence
of a key mathematical expression with that for trajectories at the
tangent bifurcation [\ref{III-ix-2}].                                     
  This fact developed into a nonlinear      
dynamical approach for rank distributions that uncovered similarities with
universality classes in critical phenomena [\ref{III-ix-1}--\ref{III-ix-8}].  
  The size-rank  
distribution of all kinds of numerical data, including Zipf law, were
seen to be obtainable from nonlinear maps close to a tangent
bifurcation, such that data samples are reproduced by their
trajectories [\ref{III-ix-1}--\ref{III-ix-7}].                                  
Remarkably, size-rank distributions with power
law decay can be reproduced by trajectories of the renormalization group fixed-point map
[\ref{III-ix-2}]. And as it turns out also for distributions with exponential  
decay for which the point of tangency shifts to infinity. More
generally, it was demonstrated that for all data source distributions
the mentioned map can be constructed and the rank distributions
determined [\ref{III-ix-7}].                                                    
That is, the stochastic and the deterministic
approaches are equivalent. This duality permits for an explicit and
quantitative distinction between size-rank -sizes of cities- and
frequency-rank -word frequencies- distributions, as the former appears
as a trajectory while the latter is a sum of positions [\ref{III-ix-6}].       
The frequency-rank distribution turns out to be the functional inverse of
the size-rank distribution [\ref{III-ix-6}].                                  
There are other surprising sets of
properties related to this topic that can be obtained from the map at
tangency. The reciprocals of the size-rank functions provide uniformly
distributed probabilities for fixed rank that lead to extensive
$q$-deformed entropies where system size is measured by sample size.
\add[r1]{Recently, we provided} [\ref{III-ix-8}]
an extension to Number Theory as we 
obtain from the fixed-point map trajectories the numbers, or
asymptotic approximations of them, for the Factorial, Natural, Prime
and Fibonacci sets. Additional features of the parallelism with
critical phenomena appear as borderline divergencies and logarithmic
corrections for the Zipf law class\change[CV]{,}{. This} in relation with the Prime numbers
and exemplified by earthquake data. The formalism links all types of
ranked distributions to a $q$-entropy.
}

\Abs{150} 

\Abs{156} 

\Abs{157} 

\Abs{172} 

\Abs{174} 

\Abs{181} 

\Abs{184} 

\Abs{188} 


\subsection{Transition to chaos as critical point.}  
\label{\myss}

\NewT{
\add[r1]{The well-known properties of families of attractors of the quadratic
map were revisited as seen through the densities (or measures) of
ensembles of trajectories. These probability densities were determined
via the Frobenius-Perron equation and trough them a novel
statistical-mechanical picture was obtained} [\ref{III-x-1}].
\add[r1]{Two families of
attractors were considered, the supercycles along the period-doubling
cascade and the Misiurewics points along the chaotic-band-splitting
cascade, together with their common accumulation point at the
transition to and out of chaos. From the densities the entropies
associated with these attractors were determined and, most remarkably,
when the collection of entropies for the two families of attractors is
viewed along the values of control parameter the familiar pattern
appears of a statistical-mechanical two-phase system separated by a
continuous phase transition, an equation of state containing a
critical point} [\ref{III-x-1}]. As we have already mentioned, the
transitions to chaos and the fluctuations at a
critical point display\change[CV]{.}{ $q$-statistical properties.}
}

\Abs{187}

\subsection{Nonlinear dynamical view of Kleiber's law.}  
\label{\myss}

\NewT{
The empirical law of Kleiber \cite{kleiber1,kleiber2}, metabolism grows as
mass to the \add[CV]{3/4th-}power, has remained an open question for the life
sciences since its formal finding in 1930. Why does this pattern hold
for more than ten orders of magnitude that consider\add[CV]{s} radically
different organisms? Even though there have appeared plausible
mechanistic models with quantitative agreements for both plants and
animals \cite{West1,West2,Maritan1}\change[CV]{. We}{, to understand this law we}
have taken the view of \add[CV]{a} universal,
very general basis, as in the case of the power laws of critical
phenomena\remove[CV]{, to understand this law}. Recently [\ref{III-xi-1}], we have
reproduced quantitatively the available data for these two kingdoms in
biology by means of a formalism that makes use of statistical
mechanics and nonlinear dynamics. We consider RG fixed-point nonlinear
iterative maps linked to an extensive $q$-entropy.  We focus on a unique
pair of universality classes that satisfy the \add[CV]{3/4th-}power law, one of them
corresponds to preferential attachment, rich gets richer, and the
other to critical processes that suppress, overcome, the cost of
motion. This conjugate pair of universality classes display, as the
real data do, curvatures of opposite signs in logarithmic scales for
small masses.
}

\Abs{189}

\subsection{Bifurcation cascades within windows.}  
\label{\myss}

\NewT{
The chaotic band attractors in the logistic map bifurcation diagram
contain an infinite collection of intervals, called windows \add[CV]{of periodicity}. These
windows contain sets of reproductions of the bifurcation diagram
itself, and again, repeatedly, similar windows within each smaller
replica, and so on, because the bifurcation diagram is a fractal
object. The intervals display power-law spacing and widths that have
been characterized a long time ago
\remove[r1]{[55zodiac, 56zodiac]}\cite{Yorke2,Capel}. There are   
other related structures in the bifurcation diagram, these are the
so-called ‘shadow’ curves and their dual ‘period’ curves
\remove[r1]{[90zodiac]}\cite{Shadow}. The shadow \add[CV]{and period} curves have mutual points of tangency and also  
self-intersections that envelop the bifurcation replicas inside the
windows. The power laws displayed by the families of windows point at
the existence of $q$-statistical-mechanical behaviors not yet
characterized, as other sets of power laws occurring in the dynamics
of quadratic and related maps have been, e.g. as in all the previous
sections. These power laws correspond to the spacing of the windows
across the bifurcation diagram, their widths, and other features
within them, as are the replicas of the main cascades in the primary
diagram. This complex structure is enveloped by the elaborate system
of shadow and period curves.  We are currently involved in the
construction of models designed to describe complex systems where the
most significant property is embedding, systems nested within systems.
}

\newpage

\section{Generalized Statistical Mechanics}\label{s-GeneralizedSM}  

\textbf{Limit of validity.}
As Constantino Tsallis himself describes [\ref{IV-i}], it was during a
coffee brake of a workshop, held in 1985 at the Instituto de F\'{i}sica
(Universidad Nacional Aut\'onoma de M\'exico) in Mexico City, that he
observed at a distance a professor talking to an attending student and
pointing out at the board the combination of symbols \change[CV]{p}{$p$}
to the power $q$
typical of the description of geometric multifractal sets. This made
him consider a generalization of the familiar entropy expression in
statistical mechanics and information theory. To his surprise, and to
many others, the generalized expression preserves to a large extent
the mathematical structure of the ordinary theory, or where there
appear variants these are obtained without much difficulty. The
initial publication of these developments \cite{Tsallis1entropy} produced
an enthusiastic response that translated into many studies and related
publications \cite{Tsallis1}, but also to skepticism
\cite{Cho1, Tsallis2entropy}.  Over the years this issue has
developed into an evolving topic, under the names of nonextensive
statistical mechanics or $q$-statistics \cite{Tsallis1} [\ref{III-iv-10}],  
which carries with it an exploration of the fundamental basis of this
branch of physics. It carries the implication, if \change[CV]{found}{appropriate} pertinent, of
a limit of validity of the firmly established Boltzmann-Gibbs (BG)
equilibrium statistical mechanics. \change[CV]{Firm}{Well-founded} answers and \add[CV]{thorough} understanding have
been awaited up to the present time while a large body of published
studies has been assembled and it is still growing.  We have developed
an interest and have carried out research in this subject since the
early stages and up to now. Our motivation has been two fold: The
quest and substantiation of concrete instances in which the BG theory
fails and the alternative theory is seen to be applicable. And
relatedly, the understanding of the breakdown, if it is indeed the
case, of the BG formalism, the underlying cause of the occurrence of a
limit of validity of an otherwise unbeaten physical theory. Our early
studies referred to our generalization of the Weierstrass random walks
[\ref{I-iv-1}, \ref{I-iv-5}] to show the connection between the RG technique and  
entropy optimization. A setting that we used to show the relationship \add[CV]{of these walks}
with the anomalous dimension in criticality \add[CV]{and found of a role for the $q$-index of $q$-statistics}
[\ref{I-iv-1}]. Then we made use
of the same random walks to discuss the transit from ordinary to
anomalous diffusion [\ref{I-iv-2}] with reference also to transport in
porous media [\ref{I-iv-3}].  Our main effort, spanning two decades, to    
understand the implications of a generalized statistical mechanics has
been focused on the field of nonlinear dynamics, specifically the
transitions to chaos in dissipative systems. This choice was and is
motivated by the fact that the chaotic attractors present in these
systems possesses the fundamental properties of BG statistical
mechanics: ergodicity and mixing, and that these two essentials
breakdown at the borderline situation, the transition attractors from
chaotic to regular behavior. The additional feature of relative
simplicity of (precise, exact) access to their properties makes \remove[CV]{these}
low-dimensional mathematical models desirable ‘numerical laboratories’
for this purpose. The entire contents of Part III of this research
report, its twelve sections, have produced concrete advances related
to the $q$-statistics issue, at the same time these studies have
revealed the areas in condensed matter physics and in complex systems
where the generalized statistical-mechanical formalism is
applicable. See the review articles in
Refs. \remove[r1]{[52,53,54]}\cite{Review1,Review2,Review3}. For an            
earlier account of the initial stages of $q$-statistics and of our
motivations to participate in the development of this subject see
Ref. [\ref{IV-i}] But it was only very recently [\ref{IV-ii}] that several    
apparently separate pieces of research came together to create an
important advance in the understanding of how the limit of validity of
the BG theory arises and in what way it is replaced by, precisely,
$q$-statistics. This fusion or amalgamation requires first a change in
perspective when considering the familiar phase space variable $x$ in
one-dimensional nonlinear iterated maps $f(x)$. Instead of seeing such
maps as systems composed of only one degree of freedom, they can be
seen to represent systems with many, infinite, degrees of freedom,
when described thermodynamically through a macroscopically observable
variable like the energy or the magnetization\add[CV]{, the map variable $x$}.
In other words, from
this viewpoint the map includes already a (generalized) thermodynamic
quantity, the result of the familiar average of (microscopic)
configurations in statistical mechanics that results in (macroscopic)
observables. Then the next step is to consider not only stationary
(equilibrium) states but also processes that lead to these final
states. We resort to the rate equations (that we have referred to in
Section II.iii) for the description of the kinetics of phase change in
condensed matter physics, such as the dissipative Landau-Ginzburg
equation \remove[r1]{[25,22entropy]}\cite{Landau1,Metiu3} [\ref{II-iii-1}]  
discussed here in Part II. We
focus on the differential equation variable, the driving force of the
equation, and its Lyapunov function
\remove[r1]{[26entropy]}\cite{Lyapunovdef}[\ref{II-iii-1}]. After this     
we show that the discrete time version of the Landau-Ginzburg equation
is a dissipative nonlinear iterated map \change[CV]{f(x)}{$f(x)$} and that the general
choice of the Renormalization Group (RG) fixed-point maps for the
three known routes to chaos capture a wide range of possible
applications (including those described in the twelve sections of Part
III). In addition to all this we demonstrate that the RG fixed point
maps and all of their trajectories have analytical closed
$q$-exponential expressions. Technically, the results of this approach
are: i) The $q$-exponential function as the fabric of the RG fixed-point
maps for the three routes to chaos. ii) Its inverse function the
$q$-logarithm, identified as the Tsallis entropy expression, that,
significantly, turns out to be the Lyapunov function of the discrete
time Landau equation. As for a brief explanation of the limit of
validity of ordinary statistical mechanics and the subsequent
incidence of Tsallis statistics: These are a consequence of the effect
of the RG fixed-point maps attractors in reducing drastically the
initial phase space, or in terms of the phenomena modeled by them, in
a severe hindrance to access initially available configurations. See
Ref. [\ref{IV-ii}] for a detailed analysis and explanation.                 

\subsection{A new statistical mechanics}     
\label{\myss}
\add[CV]{``If fully chaotic dynamics lays the foundation for statistical
  mechanics, the question arises: What happens when dynamics is only
  incipiently chaotic? What are the limits of validity of traditional
  statistical mechanics? Is there any generalization of the existing
  theory?''}

\Abs{190}

\subsection{How, why and when is \emph{q}-statistics pertinent}
\label{\myss}
\add[CV]{``The RG fixed-point maps can be seen as discrete-time versions
  of the Landau kinetic equation and they are shown to be associated
  with a Lyapunov function given by the Tsallis entropy. The monotonic
  iteration time evolution of this function obeys $q$-statistics and
  displays $q$-exponential partition function configuration weights.''}

\Abs{191}

\newpage


\section*{Summary and outlook}
\addcontentsline{toc}{section}{Summary and outlook}

  We have summarized a large catalogue of contributions on stochastic
  processes, statistical physics, and nonlinear dynamics with
  applications to condensed-matter physics and to complex system
  phenomena common to various disciplines. The contributions were
  created and elaborated basically in Mexico (except for sabbatical
  periods and other causes) over the last fifty years. The studies
  have been grouped into \change[CV]{twenty-seven}{thirtytwo} sets and these,
  in turn, into
  \change[CV]{three}{four} parts. The order of presentation is not quite chronological as
  it records reappearances due to unanticipated opportunities to
  crosslink methodologies that led to interesting analogies. We refer
  to some of these below, as well as to a few other present ongoing
  investigations.
  
  
  \emph{Stochastic Processes.} Our early works on random walks in the 70's
  and early 80's  [\ref{I-i}]
  became most useful in the late 90's,
  when, via the generalization of the Weierstrass walk, it became
  apparent that this famous walk is the nontrivial fixed point of the
  Renormalization Group transformation devised for this case [\ref{I-iv}].
  More importantly, this fact served as the means to show that
  entropy optimization had remained an unseen but significant tool for
  this method. Later, [\ref{II-iv-5}],
  we showed that continuous time
  random walks could be combined with the nonlinear dynamics of
  deterministic intermittency, to provide an unforeseen view of
  anomalous diffusion. Another random walk process advance, still to
  be fully developed, is to recast our results for sums of positions
  at chaotic-band attractors
  [\ref{s-DensityFT}, \ref{II-iv-1}-\ref{II-iv-5}]
  in terms of correlated
  walks. Lastly, we mention our work on renewal processes
  [\ref{I-v}],
  where the expressions
  for the event probabilities, those for
  their generating functions, and those for their Laplace transforms,
  comply with the properties of customary partition functions and
  ensemble equivalence of equilibrium statistical mechanics. This
  establishes a strict analogy between each thermal system model and a
  renewal process. Since the ideal gas maps into the case of
  independent events, each particle-interacting model describes a
  renewal process with correlated events
  [\ref{I-vi}].
  
  \emph{Density Functional Theory.} The order parameter spatial shape of the
dominant fluctuation at a critical point can be explored via density
functional theory just as it is the case for equilibrium
inhomogeneities, interfacial profiles or confined fluids. The square
gradient approximation is suitable for this purpose
[\ref{III-i}].
The difference with the standard applications is that the fluctuation is
unstable and has necessarily a finite lifetime. Its stability can be
studied via the 2nd variation of the density functional and its time
evolution via the corresponding Landau differential equation
[\ref{III-i-4}].
This procedure makes contact with nonlinear dynamical problem,
that of intermittency close to a tangent bifurcation
\cite{Schuster1}. Another interesting crosslink with the renewal processes
studied in Part I is to consider correlated renewal events via Widom's
insertion method \cite{Widom2}. The translation of a showcase thermal
system, the so-called Hamiltonian Mean Field Model leads to a renewal
process that undergoes a phase transition along time [\ref{I-vi}].
Yet another instance involves the bifurcation pattern we found in our
treatment of hard-core models under the Bethe approximation
[\ref{II-ii-3}],
consisting of the instability of the uniform-fluid solution
at the highest density, the close-packing density. There, nontrivial
oscillatory solid-like solutions appear for the 1st time. A robust
analogous behavior can be found in the nonlinear dynamical model for
epidemic spreading \cite{Nowak1}. This opens the possibility of treating
such complex system problems, like obesity \cite{Camacho1}, with Widom's
particle-addition statistical-mechanical technique.

  \emph{Nonlinear Dynamics.} A number of ongoing studies associated with our
research lines on nonlinear dynamics have been just described and
explained \cite{Review3}. In addition to them we mention three sets of the
most recent investigations. First, we explore new access to
statistical-mechanical viewpoints from the known dynamical behavior of
the quadratic map attractors by considering the alternative
description via probability densities of ensembles of trajectories
instead of the trajectories themselves \cite{Alvaro1}. Through this we
observe the equivalence between the two main paradigms for the
understanding of complex systems in recent decades: edge of chaos and
criticality. Second, for some time now we have represented the
(master) trajectory at the period-doubling onset of chaos in
logarithmic scales [\ref{III-iv-3}].
This representation reveals an
infinite family of straight lines, or interlaced power laws, that have
led to many fruitful interpretations and analytical
statistical-mechanical results \cite{Review1,Review2,Review3}. This
feature has become now a pathway to demonstrate a common link between
all renormalization-group fixed-point maps at the transitions to
chaos, intermittency, period doubling and quasiperiodicity. Third, the
control parameter gaps that interrupt the chaotic attractor intervals
in the bifurcation diagram of the quadratic map contain infinite
families of reproductions of the bifurcation diagram itself, the gaps
display power-law spacing and widths \cite{Yorke2}. These features are
caricature building blocks for current modeling of nested complex
systems.

The above brief commentaries serve as indications for some feasible
short term extensions of the long footpath described here.
\add[CV]{But also, given the potentially significative general results
  described in Section IV, there is a wider road to enter and engage in
  \change[CV]{in to study further, and in various aspects,}{in:} the limit of validity
  of ordinary statistical mechanics and the consequent prevalence of a
  generalized statistical mechanics.}

\newpage

\section*{Acknowledgments}
\addcontentsline{toc}{section}{Acknowledgments}

AR is profoundly thankful to all his collaborators over the years and
deeply appreciative to all colleagues with whom valuable discussions
took place. Support is acknowledged from IN106120-PAPIIT-DGAPA-UNAM
and 39572-Ciencia-de-Frontera-CONACyT.

\newpage

\renewcommand\refname{References}

\newpage               

\section*{Figure Descriptions}
\phantomsection\addcontentsline{toc}{section}{Figure Descriptions}

\quad \ Figure 1. Random walks. The top panels illustrate trajectories for random walks with steps of varying lengths for which the balance from short to long step lengths is governed by a parameter $\mu$. The middle panels show the probability distribution $P_{n} (l)$ for a walker to be at a position $l$ after $n$ steps for different values of $n$. Two different values of $\mu$ were chosen above and below a critical threshold $\mu_c=2$. The bottom panels show the same probabilities in the middle panels, in one case in logarithmic scale for $P$ and quadratic scale for $l$, and in the other case in logarithmic scales for both $P$ and $l$, to corroborate the evolution towards Gaussian and Levy stationary distributions in the limit $n\rightarrow \infty$ for the chosen values of $\mu$. Studies related to these figure panels are found in Section I iv.

Figure 2. Wetting phenomena and complex fluid models. The top left panel shows the density or composition inhomogeneity that becomes a line of tension in macroscopic scales. The top right panel illustrates drops of different liquid phases on a flat solid surface that form lines of tension when three different interfaces meet. The left middle panel shows a global phase diagram for wetting and prewetting interfacial transitions and where different types of multicritical points occur. Sections II iv and II x refer to studies related to these figure panels. The right middle panel illustrates lattice model configurations for different types of bifunctional molecules that are placed along lattice bonds with certain rules. The bottom panels show phase diagrams for complex fluids obtained for these types of lattice models. Section II v refers to studies related to these figure panels.

Figure 3. Reentrant solubility and High-$T_c$ superconductors phase behavior. The top panels show the phenomenon of reentrant solubility that involves two different critical miscibility points. The plots correspond to micellar solution lattice models that map into the Ising magnet model. The asymmetric case involves anomalous criticality. Section II vi refers to studies related to these figure panels. The bottom panels show model phase diagrams for sublattice ordering and structural transitions and comparison with experimental counterpart in High-$T_c$ copper-oxide superconductors. Section II xii refers to studies related to these figure panels.

Figure 4. Dynamics at the transition to chaos. The top panels show anomalous properties at the transition to chaos in quadratic maps. In the left we see the expansion rate that implies vanishing Lyapunov exponent but rich behavior for the sensitivity to initial conditions. The right panel describes the interwoven power law structure of a trajectory inside the atractor. Section III iv refers to studies related to these figure panels. The bottom panels show dynamical properties of an ensemble of trajectories running towards periodic attractors. In the left is seen how a family of gaps is formed sequentially by initially uniformly distributed trajectories. In the right is shown the power-law decay with logarithmic oscillations of phase space occupancy, where recapitulation is observed along the period-doubling family of attractors. Section III vi refers to studies related to these figure panels. 

Figure 5. Modeling of complex systems via nonlinear dynamics. The left top panel represents in color the values for the Lyapunov exponent for the discrete time symmetric two-strategy cooperation games, including the prisoner's dilemma. Section III vii refers to studies related to this figure panel. The right top panel illustrates maps close to a tangent bifurcation with properties equivalent to a model for localized to extended transport properties via scattering media. Section III iii refers to studies related to these figure panels.  The middle panel describes the development of a fractal function that represents sums of positions for ensembles of trajectories evolving under the action of period doubling attractors. Section III v refers to studies related to this figure panel. The bottom panel illustrates the period doubling and chaotic band splitting attractor cascades as seen via (Horizontal Visibility) networks. Sections III viii refers to studies related to this figure panel.

\begin{figure}[!htb]
  \centering
  \includegraphics[width=0.8\linewidth]{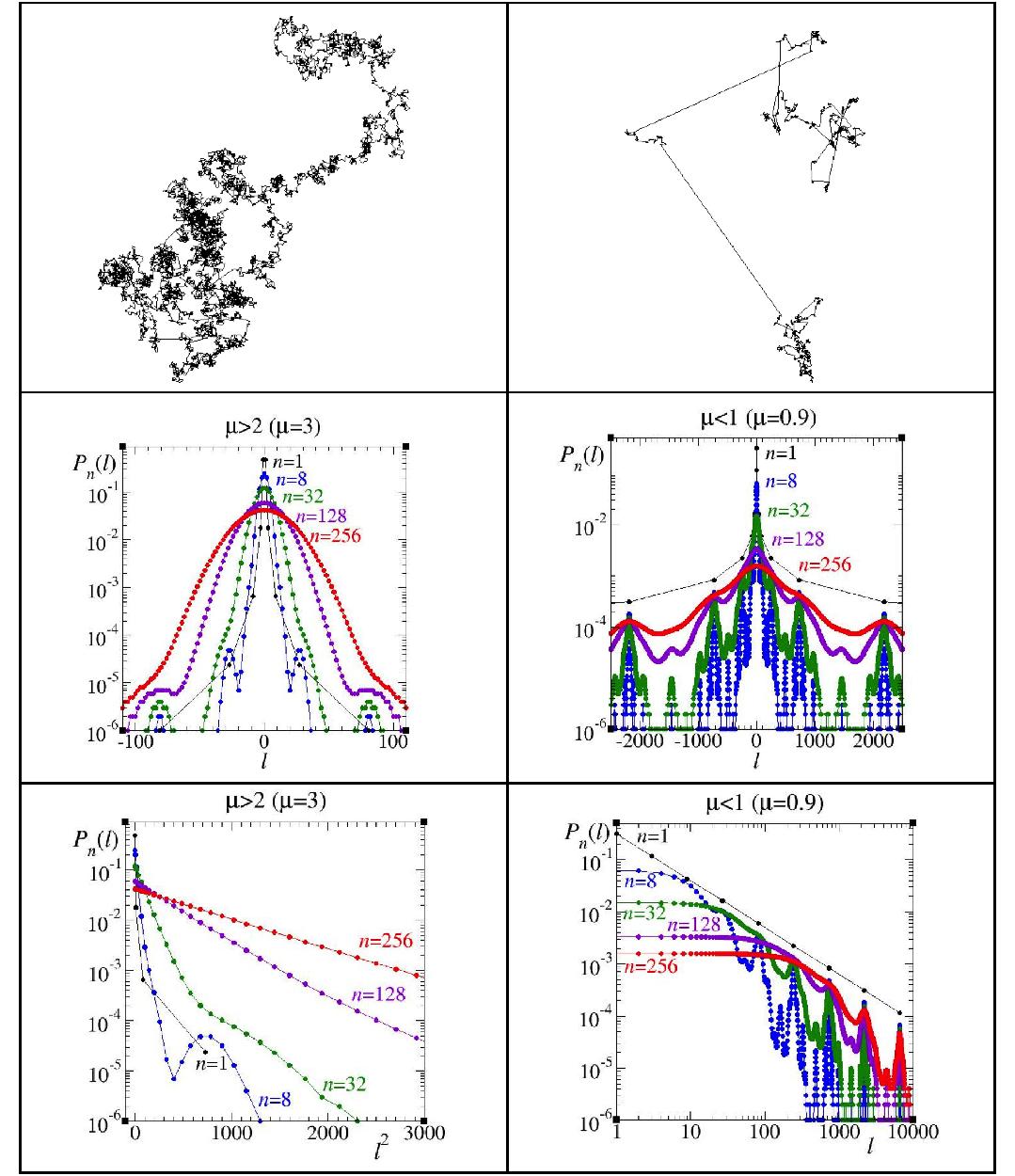} 
  \caption{\footnotesize }
  \label{Fig1}
\end{figure}

\newpage

\begin{figure}[!htb]
  \centering
  \includegraphics[width=0.8\linewidth]{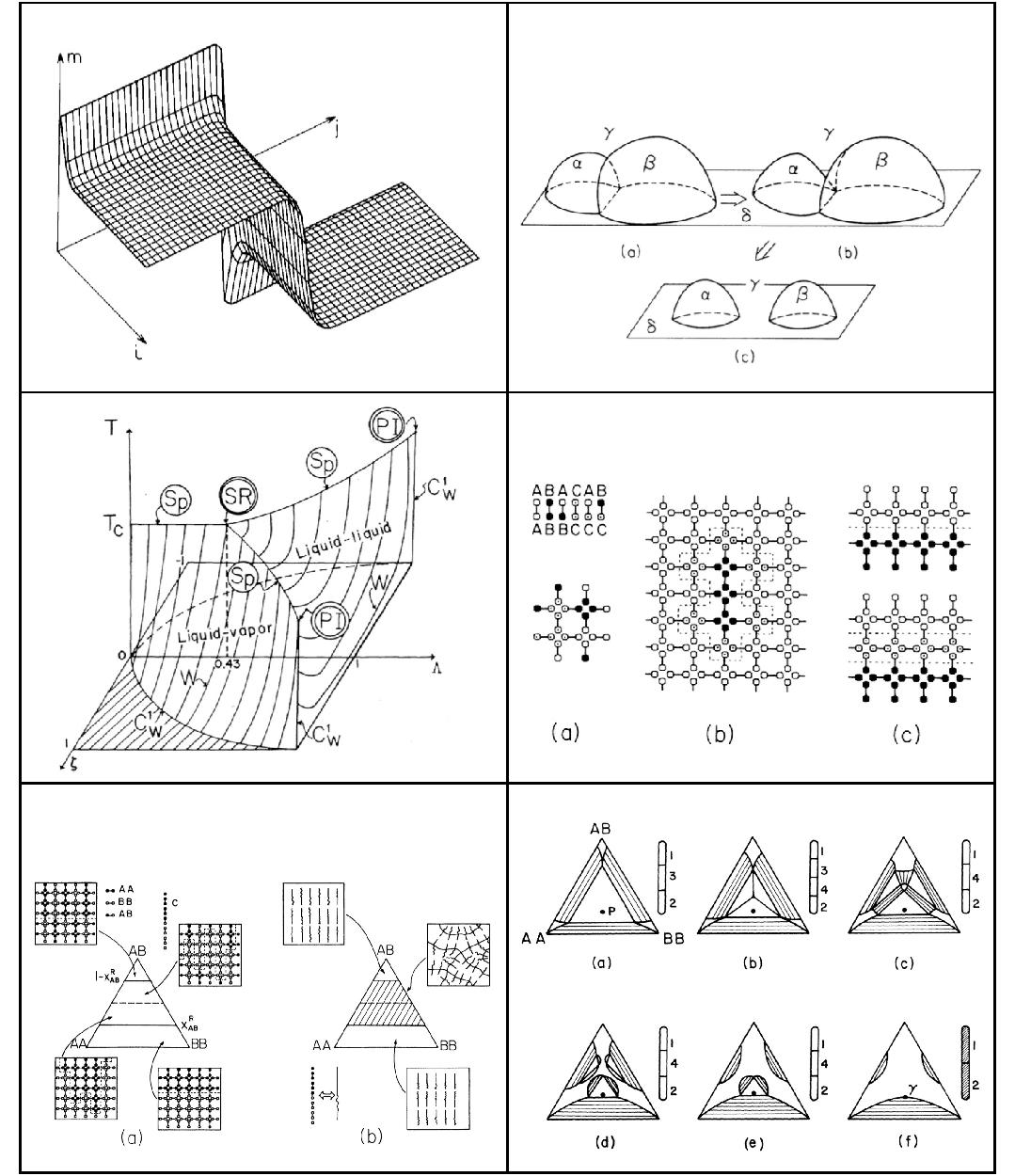} 
  \caption{\footnotesize }
  \label{Fig2A}
\end{figure}

\newpage

\begin{figure}[!htb]
  \centering
  \includegraphics[width=0.8\linewidth]{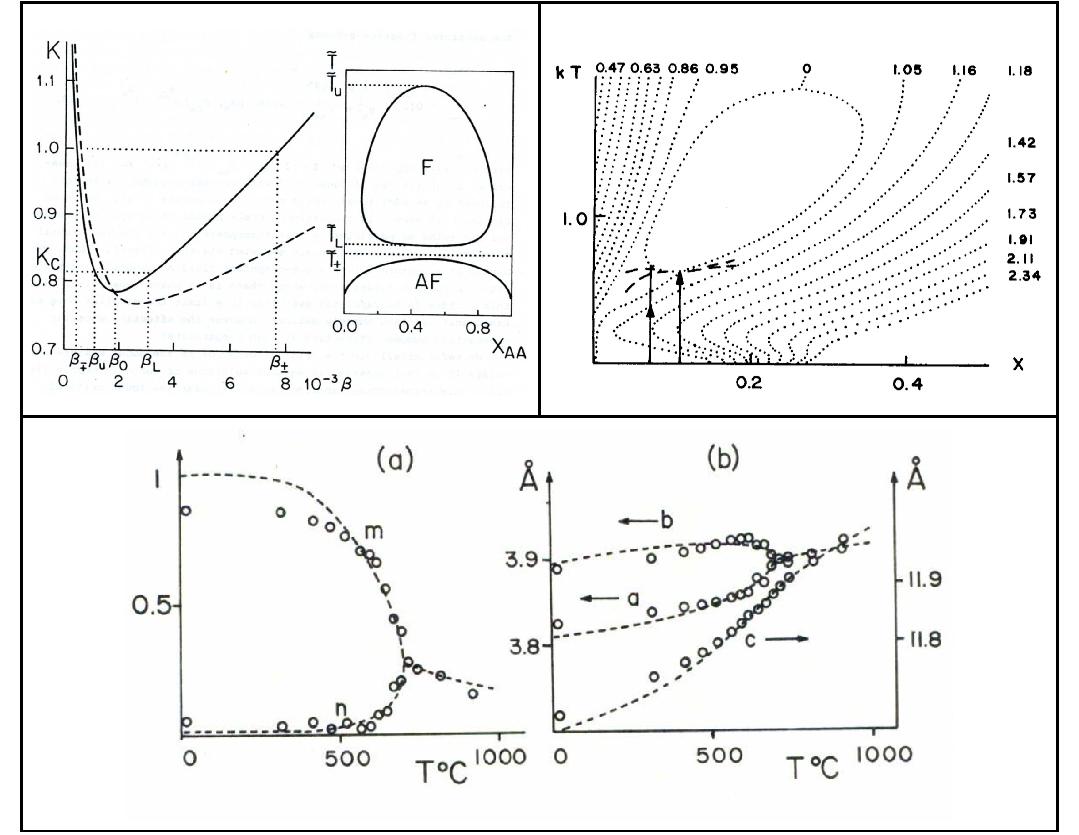} 
  \caption{\footnotesize }
  \label{Fig2B}
\end{figure}

\newpage

\begin{figure}[!htb]
  \centering
  \includegraphics[width=0.8\linewidth]{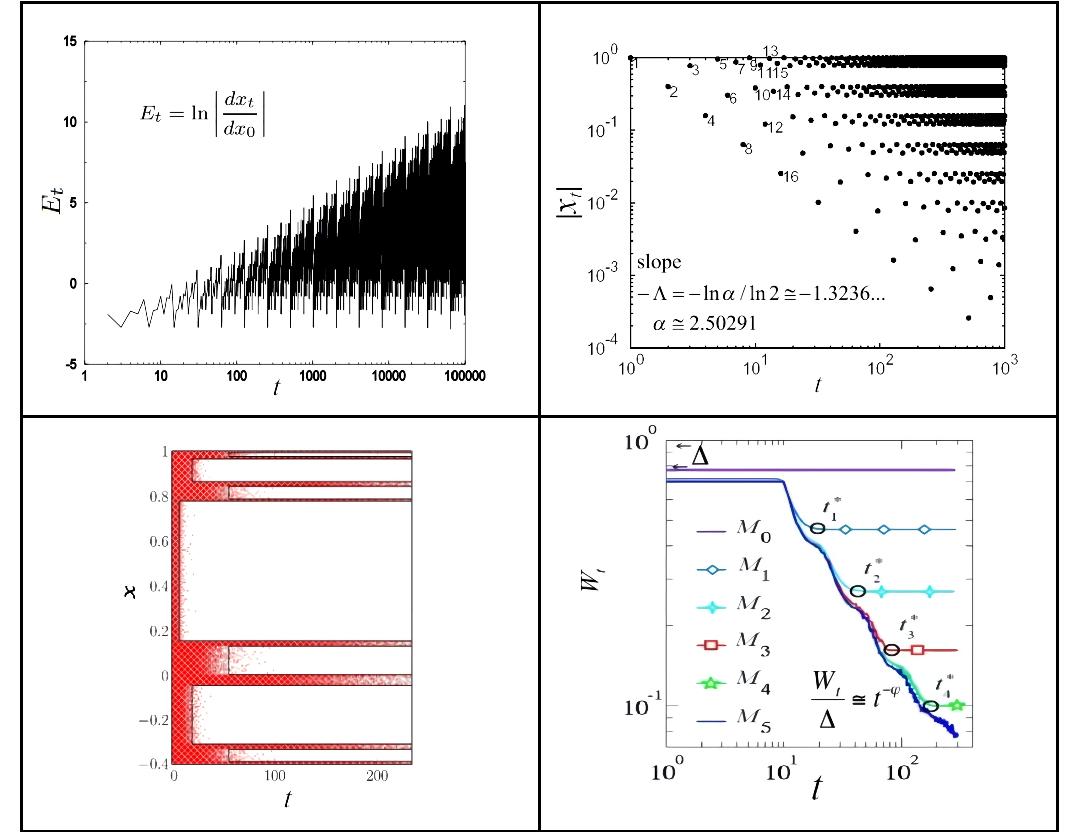} 
  \caption{\footnotesize }
  \label{Fig3A}
\end{figure}

\newpage

\begin{figure}[!htb]
  \centering
  \includegraphics[width=0.8\linewidth]{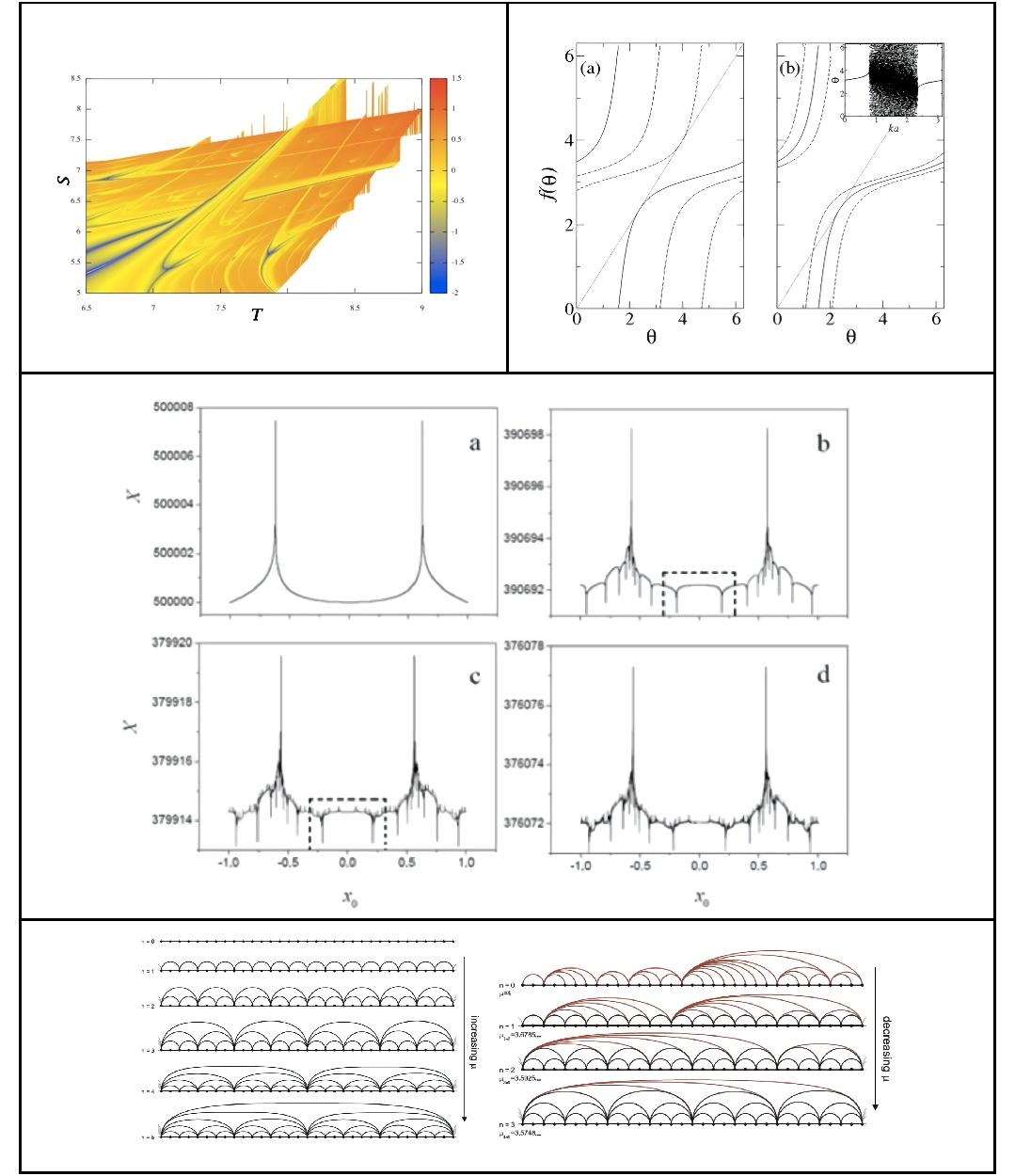} 
  \caption{\footnotesize }
  \label{Fig3B}
\end{figure}

\end{document}